\documentclass[twocolumn]{aastex7}
\usepackage[T1]{fontenc} % for Icelandic Characters
\usepackage[utf8]{inputenc} % for Icelandic Characters 

\newcommand\tess{TESS}
\newcommand\gaia{\textit{Gaia}}
\newcommand\kms{$\textrm{km~s}^{-1}$}
\newcommand\ms{$\textrm{m~s}^{-1}$}
\newcommand\cms{$\textrm{cm~s}^{-1}$}
\newcommand\gcmcubed{$\textrm{g~cm}^{-3}$}

\newcommand\teff{T$_{\rm{eff}}$}
\newcommand\vsini{$v$~sin~$i$}

\newcommand{\unit}[1]{\ensuremath{\, \mathrm{#1}}} %Remove if necessary
\newcommand\earthmass{$M_{\oplus}$}
\newcommand\earthradius{$R_{\oplus}$}
\newcommand\solmass{$M_{\odot}$}

\received{21 June, 2025}
\revised{23 July, 2025}
\accepted{28 July, 2025}
\submitjournal{AJ}

\begin{document}

\title{\textit{Searching for GEMS:}  TOI-7149~b an Inflated Giant Planet causing a 12 \%~Transit of a Fully Convective M-dwarf}

\author[0000-0001-8401-4300]{Shubham Kanodia}
\affiliation{Carnegie Science Earth and Planets Laboratory, 5241 Broad Branch Road, NW, Washington, DC 20015, USA}
\email[show]{skanodia@carnegiescience.edu}  

\author[0000-0003-4835-0619]{Caleb I. Ca\~nas}
\altaffiliation{NASA Postdoctoral Fellow}
\affiliation{NASA Goddard Space Flight Center, 8800 Greenbelt Road, Greenbelt, MD 20771, USA }
\email[]{c.canas@nasa.gov}

\author[0000-0001-9596-7983]{Suvrath Mahadevan}
\affil{Department of Astronomy \& Astrophysics, 525 Davey Laboratory, The Pennsylvania State University, University Park, PA 16802, USA}
\affil{Center for Exoplanets and Habitable Worlds, 525 Davey Laboratory, The Pennsylvania State University, University Park, PA 16802, USA}
\email[]{suvrath@astro.psu.edu}

\author[0000-0002-9082-6337]{Andrea S.J.\ Lin} %%%
\affiliation{Department of Astronomy, California Institute of Technology, 1200 E California Blvd, Pasadena, CA 91125, USA}
\email{asjlin@caltech.edu}

%%%% RBO %%%%
\author[0000-0002-4475-4176]{Henry A. Kobulnicky}
\affil{Department of Physics \& Astronomy, University of Wyoming, Laramie, WY 82070, USA}
\email[]{chipk@uwyo.edu}  

\author[0009-0009-6263-0490]{Ian Karfs} %%%
\affil{Department of Physics \& Astronomy, University of Wyoming, Laramie, WY 82070, USA}
\email[]{ikarfs@uwyo.edu}  

\author[]{Alexina Birkholz} %%%
\affil{Department of Physics \& Astronomy, University of Wyoming, Laramie, WY 82070, USA}
\email[]{abirkho1@uwyo.edu}  

\author[0000-0002-0048-2586]{Andrew Monson} %%%
\affil{Steward Observatory, The University of Arizona, 933 N.\ Cherry Avenue, Tucson, AZ 85721, USA}
\email{andymonson@arizona.edu}

\author[0000-0002-5463-9980]{Arvind F.\ Gupta} %%%
\affil{U.S. National Science Foundation National Optical-Infrared Astronomy Research Laboratory, 950 N. Cherry Ave., Tucson, AZ 85719, USA}
\email{arvind.gupta@noirlab.edu}

\author[]{Mark Everett} %%%
\affil{U.S. National Science Foundation National Optical-Infrared Astronomy Research Laboratory, 950 N. Cherry Ave., Tucson, AZ 85719, USA}
\email{mark.everett@noirlab.edu}

\author[0009-0009-4977-1010]{Michael Rodruck} %%%
\affil{Department of Physics, Engineering, and Astrophysics, Randolph-Macon College, Ashland, VA 23005, USA}
\email{mrodruck@gmail.com} 

\author[0000-0001-9816-0878]{Rowen I. Glusman} %%%
\affil{Anton Pannekoek Institute for Astronomy, University of Amsterdam, Science Park 904, 1098 XH Amsterdam, The Netherlands} 
\email{rowen.glusman@student.uva.nl}

\author[0000-0002-7127-7643]{Te Han} %%%
\affil{Department of Physics \& Astronomy, The University of California, Irvine, Irvine, CA 92697, USA}
\email{teh2@uci.edu}

%HPF
\author[0000-0001-9662-3496]{William D. Cochran} %%%
\affil{McDonald Observatory and Department of Astronomy, The University of Texas at Austin}
\affil{Center for Planetary Systems Habitability, The University of Texas at Austin}
\email{wdc@astro.as.utexas.edu}

%HPF
\author[0000-0003-4384-7220]{Chad F.\ Bender} %%%
\affil{Steward Observatory, The University of Arizona, 933 N.\ Cherry Avenue, Tucson, AZ 85721, USA}
\email{cbender@arizona.edu}

%HPF
\author[0000-0002-2144-0764]{Scott A. Diddams} %%%
\affil{Electrical, Computer \& Energy Engineering, University of Colorado, 1111 Engineering Dr.,  Boulder, CO 80309, USA}
\affil{Department of Physics, University of Colorado, 2000 Colorado Avenue, Boulder, CO 80309, USA}
\email[]{scott.diddams@colorado.edu}  

%HPF
\author[0000-0001-9626-0613]{Daniel Krolikowski} %%%
\affil{Steward Observatory, The University of Arizona, 933 N. Cherry Ave, Tucson, AZ 85721, USA}
\email{krolikowski@arizona.edu}

% HPF
\author[0000-0003-1312-9391]{Samuel Halverson} %%%
\affiliation{Jet Propulsion Laboratory, California Institute of Technology, 4800 Oak Grove Drive, Pasadena, California 91109}
\email{samuel.halverson@jpl.nasa.gov}

%HPF/NEID
\author[0000-0002-2990-7613]{Jessica Libby-Roberts} %%%
\affil{Department of Astronomy \& Astrophysics, 525 Davey Laboratory, The Pennsylvania State University, University Park, PA 16802, USA}
\affil{Center for Exoplanets and Habitable Worlds, 525 Davey Laboratory, The Pennsylvania State University, University Park, PA 16802, USA}
\email{jer5346@psu.edu}

%HPF 
\author[0000-0001-8720-5612]{Joe P.\ Ninan} %%%
\affil{Department of Astronomy and Astrophysics, Tata Institute of Fundamental Research, Homi Bhabha Road, Colaba, Mumbai 400005, India}
\email{joe.ninan@tifr.res.in}

%HPF 
\author[0000-0003-0149-9678]{Paul Robertson} %%%
\affiliation{Department of Physics \& Astronomy, The University of California, Irvine, Irvine, CA 92697, USA}
\email{paul.robertson@uci.edu}

\author[0000-0001-8127-5775]{Arpita Roy} %%%
\affiliation{Astrophysics \& Space Institute, Schmidt Sciences, New York, NY 10011, USA}
\email{arpita308@gmail.com}

% HPF
\author[0000-0002-4046-987X]{Christian Schwab} %%%
\affil{School of Mathematical and Physical Sciences, Macquarie University, Balaclava Road, North Ryde, NSW 2109, Australia}
\email{mail.chris.schwab@gmail.com}

%HPF
\author[0000-0001-7409-5688]{Guðmundur Stef\'ansson} %%%
\affil{Anton Pannekoek Institute for Astronomy, University of Amsterdam, Science Park 904, 1098 XH Amsterdam, The Netherlands} 
\email{g.k.stefansson@uva.nl}

\correspondingauthor{Shubham Kanodia}
\email{skanodia@carnegiescience.edu}

\begin{abstract}
We describe the discovery and characterization of TOI-7149~b, a 0.705 $\pm$ 0.075 $M_J$, 1.18 $\pm$ 0.045 $R_J$ gas giant on a $\sim 2.65$ day period orbit transiting an M4V star with a mass of 0.344 $\pm$ 0.030~\solmass{} and an effective temperature of 3363 $\pm$ 59 K. The planet was first discovered using NASA's TESS mission, which we confirmed using a combination of ground-based photometry, radial velocities, and speckle imaging. The planet has one of the deepest transits of all known main-sequence planet hosts at $\sim$ 12\% ($R_p/R_\star\sim 0.33$). Pushing the bounds of previous discoveries of \underline{G}iant \underline{E}xoplanets around \underline{M}-dwarf \underline{S}tars (GEMS), TOI-7149 is one of the lowest mass M-dwarfs to host a transiting giant planet. We compare the sample of transiting GEMS to stars within 200 pc with a Gaia colour magnitude diagram (CMD) and find that the GEMS hosts are likely to be high metallicity stars. We also  analyze the sample of transiting giant planets using the non-parametric \texttt{MRExo} framework to compare the bulk density of warm Jupiters across stellar masses. We confirm our previous result that transiting Jupiters around early M-dwarfs have similar masses and densities to warm Jupiters around FGK stars, and extend this to mid M-dwarfs, thereby suggesting a potential commonality in their formation mechanisms.
\end{abstract}


\section{Introduction} \label{sec:intro}

TESS' all sky coverage includes millions of bright nearby stars \citep{ricker_transiting_2014}, particularly M-dwarfs. These stars form the majority of the stars in the Galaxy \citep{henry_solar_2006, reyle_10_2021}, and have been well-studied for their abundance of terrestrial planets, both from the transit technique with Kepler/K2 \citep[][etc.]{dressing_occurrence_2013, hardegree-ullman_kepler_2019, hsu_occurrence_2020} and radial velocities \citep[RV; ][etc.]{bonfils_harps_2013, sabotta_carmenes_2021}. However the study of giant planets around M-dwarfs has historically been facilitated by minimum mass measurements of non-transiting RV detections \citep{endl_exploring_2006, johnson_giant_2010, bonfils_harps_2013, maldonado_connecting_2019, sabotta_carmenes_2021, schlecker_rv-detected_2022, pass_mid--late_2023}, the planetary natures of which are uncertain due to the $\sin{i}$ degeneracy. This has been changing recently due to detections of giant exoplanets around M-dwarf stars (GEMS) from TESS \citep[e.g.,][]{canas_warm_2020, kanodia_toi-5205b_2023, hobson_toi-3235_2023,  bernabo_searching_2024, fernandes_searching_2025} that have enabled statistical analysis of the bulk properties of these planets \citep{muller_bulk_2024, kanodia_transiting_2024}, their host star metallicity dependence \citep{gan_metallicity_2024}, their atmospheres \citep{canas_gems_2025}, and their occurrence rates \citep{gan_occurrence_2023, bryant_occurrence_2023}. However, this sample is still limited, and in particular, is biased towards early M-dwarf hosts. Thus we have started the \textit{Searching for GEMS} survey \citep{kanodia_searching_2024} to help mitigate this.

TIC-459323923.01 was identified as a planet candidate during development of our custom \texttt{TESS-miner python} package (Glusman et al. in prep.), which was developed to identify transiting GEMS in a volume-limited 200 pc sample of $\sim$ 1 million M-dwarfs observed by TESS  as part of the \textit{Searching for GEMS} survey \citep{kanodia_searching_2024}. It was identified by our pipeline in May 2024, after which we began ground-based observations. TIC-459323923.01 was subsequently identified as a TESS object of interest (TOI) 7149.01 by the Faint-star Search \citep{kunimoto_tess_2022} using the Quick Look Pipeline (QLP) algorithm developed by \cite{huang_photometry_2020} in October 2024. We adopt TOI-7149 as the host star name for this manuscript and discuss our observations in Section \ref{sec:observations}, which include TESS photometry as well as ground-based photometry from Red Buttes Observatory, the 200-inch Hale Telescope at Palomar, and the Three-hundred MilliMeter Telescope and Las Campanas Remote Observatory at Las Campanas. We then describe our spectroscopic follow-up using the Habitable-zone Planet Finder (HPF). In Section \ref{sec:stellar} we discuss the estimated stellar characteristics, and the procedure followed to fit the observations and obtain planetary parameters in Section \ref{sec:joint}. In Section \ref{sec:discussion}, we contextualize TOI-7149~b in the landscape of other transiting GEMS, as well as discuss sample-level trends seen for transiting giant planets across the stellar mass axis, before summarizing our results and concluding in Section \ref{sec:conclusion}.

\section{Observations}\label{sec:observations}

\begin{figure*}[!t]
\begin{center}
\fig{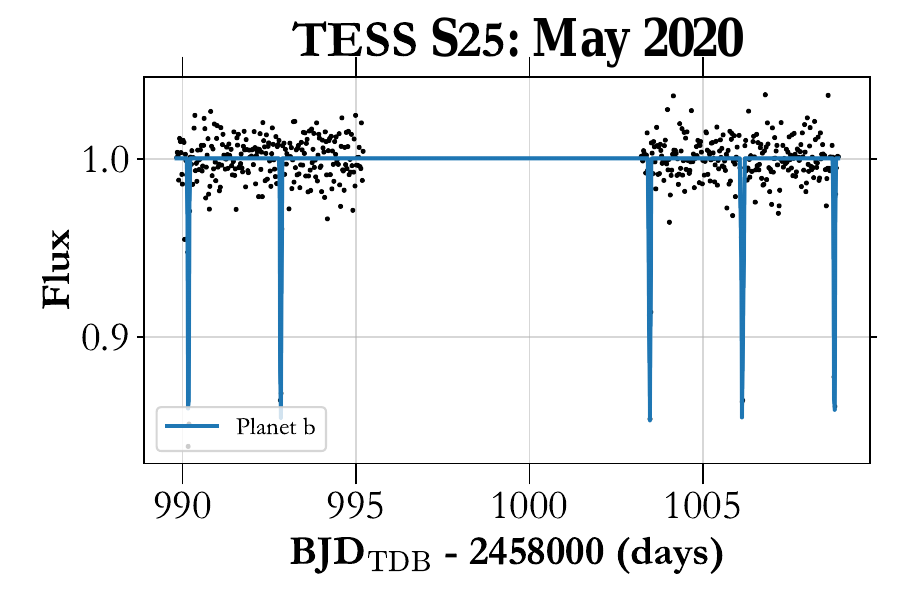}{0.48\textwidth}{}
\fig{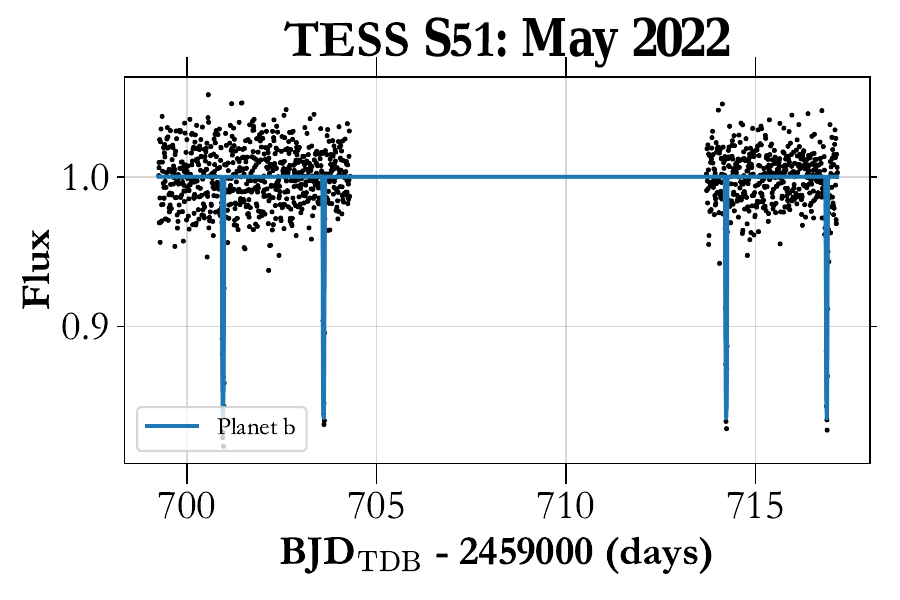}{0.48\textwidth}{}
\fig{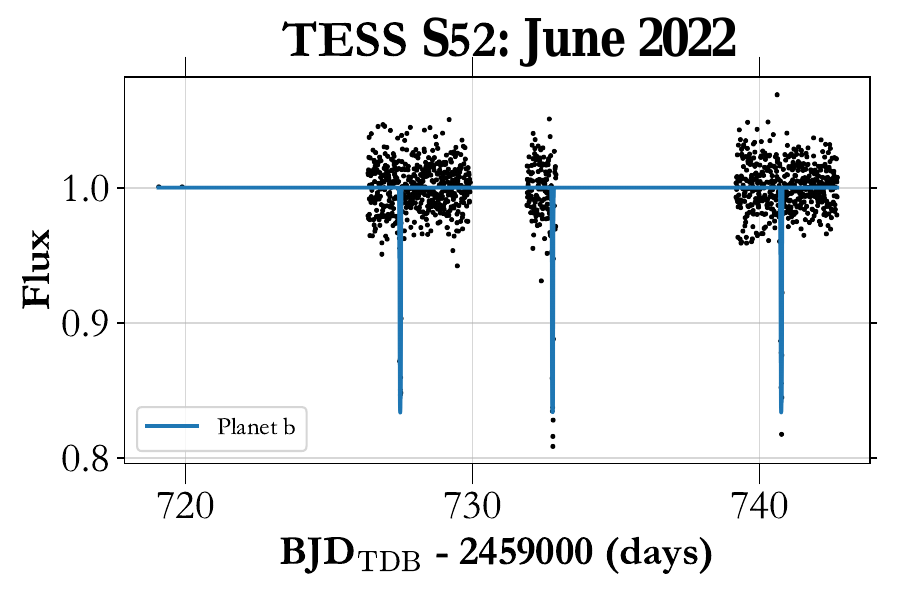}{0.48\textwidth}{}
\fig{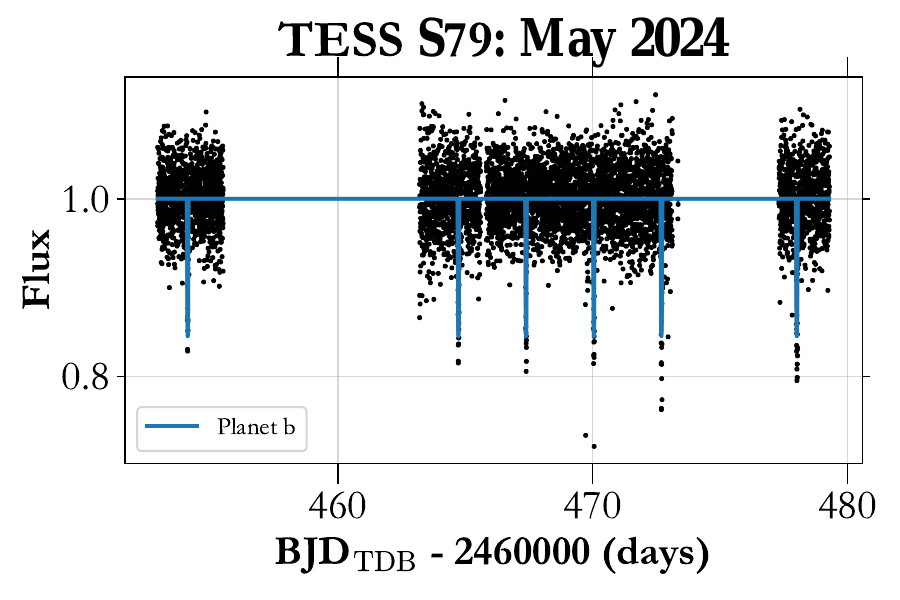}{0.48\textwidth}{}
\end{center}
% \vspace{-2 cm}
\caption{TESS photometry across sectors 25, 51, 52 and 79, with the best-fit model shown in blue. }\label{fig:tess_lc}
\end{figure*}

\subsection{TESS}\label{sec:TESS}

\begin{figure}[!t] 
\centering
\includegraphics[width=0.45\textwidth]{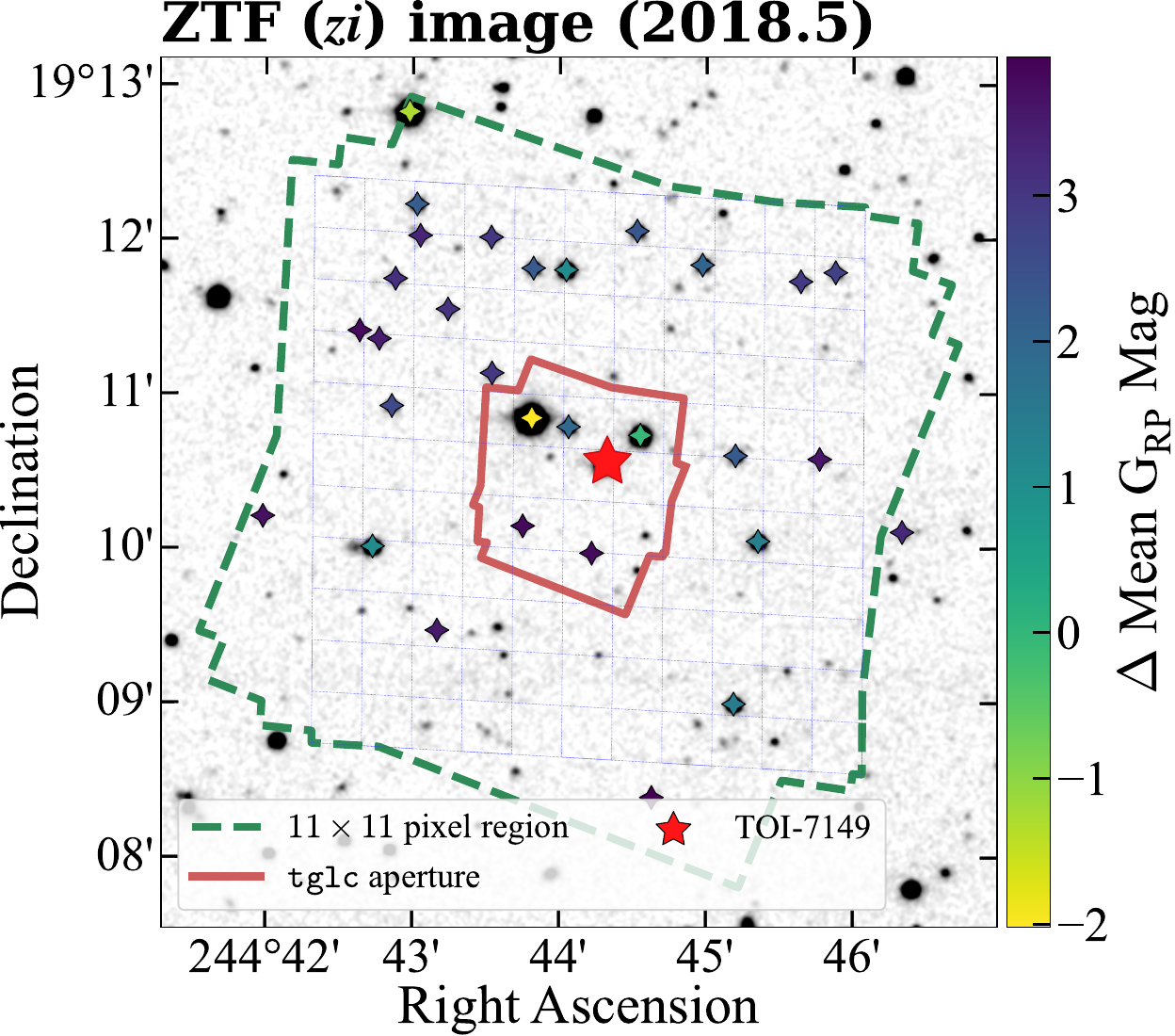}
\caption{Overlaying the \texttt{tglc} $3 \times 3$ pixel aperture for all sectors (red) on a $11 \times 11$ TESS pixel footprint for all sectors. The blue grid is the $11\times11$ footprint for Sector 79. TOI-7149 is shown as a red star. Stars with $|\Delta G_{RP}|<4$ contained in the footprint are coloured by their $\Delta$ G$_{RP}$ mag, showing the crowded field. The background image is from ZTF $zi$ around 2018.5 \citep{masci_zwicky_2019}.} \label{fig:tesspixel}
\end{figure}

TESS observed TOI-7149 across five sectors (25, 51, 52, 78, 79) with a baseline spanning $\sim4$ years. Sector 78 is excluded from further analysis due to excessive stray and background light contamination over the majority of the light curve. Of the remaining sectors, TOI-7149 was observed in Sector 25 in May 2020 with an 1800~s cadence, in Sectors $51-52$ from May--June 2022 at 600~s cadence, and in Sector 79 in June 2024 with a 200~s cadence. 

We extracted the light curves from the TESS full-frame images (FFIs) using the TESS-Gaia Light Curve (\texttt{tglc}\footnote{\url{https://github.com/TeHanHunter/TESS_Gaia_Light_Curve}}) pipeline \citep{han_tess-gaia_2023}. \texttt{tglc} models the TESS point spread function (PSF) along with Gaia astrometry to estimate and correct for contamination in TESS photometry from nearby stars. 

The \texttt{tglc} photometry for Sectors 25, 51, and 52 was extracted as part of the light curve extraction procedure for the survey (Glusman et al. in prep.), where we used a $150\times150$ pixel cutout to estimate the empirical PSF needed to correct background contamination. For Sector 79, we used a $91\times91$ pixel cutout. As part of this, we also perform a sigma-clip to the aperture flux from \texttt{tglc} and detrend the clipped time series using \texttt{wotan} \citep{hippke_wotan_2019}. We first employ a `cosine' kernel with a kernel length of 10 days using \texttt{wotan} to remove low-frequency structure from the light curve. We then run a box-least squares \citep[BLS;][]{kovacs_box-fitting_2002} periodogram over a period range of 1.0 to 12.0 days, and mask out the identified transits, before running a second detrending with \texttt{wotan}, this time with a higher frequency kernel of 0.5 days. These double-detrended light curves detect the planetary transit with a period of $\sim 2.65$ days, and are used for subsequent analysis and are shown in \autoref{fig:tess_lc}.

The closest background star, TIC-459323925 (Gaia DR3~1200751810900167040), is $\sim$ 0.6 mag brighter in Gaia $G$ band, $\sim0.1$ mag brighter in TESS $T$ band, and $\sim$ 17\arcsec{} away from TOI-7149 (\autoref{fig:tesspixel}). This background star causes considerable dilution in TESS photometry, which necessitates the use of the \texttt{tglc} reduction to correct dilution. This contamination is also evinced by the TIC contamination ratio \citep{stassun_revised_2019} of 0.37 on ExoFOP\footnote{\url{https://exofop.ipac.caltech.edu/tess/target.php?id=459323923}}, which should be 0 when there is no light from additional sources in the aperture. While we use  \texttt{tglc} to obtain a first-order dilution correction, we constrain the dilution further using ground-based observations that spatially resolve the background stars. 

\begin{figure}[!t] 
\centering
\includegraphics[width=0.4\textwidth]{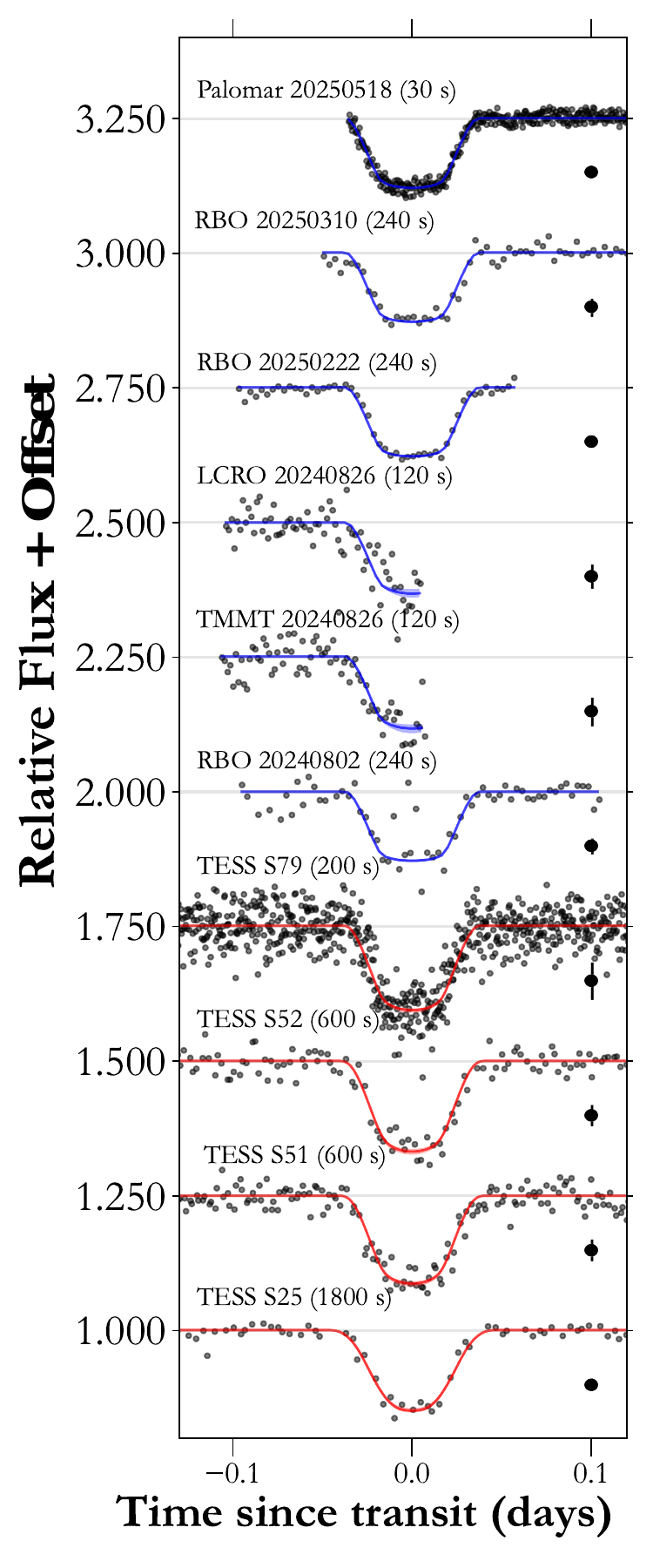}
\vspace{-0.5cm}
\caption{Photometric observations for TOI-7149~b phase-folded to the best fit orbital period; each dataset is separated by 0.25. The black points show the detrended data, while the model is shown as a solid, coloured line along with the 1-$\sigma$ confidence intervals as translucent bands. We include the representative median statistical uncertainty at x = 0.1. Ground-based observations are used to measure the transit depth and ephemeris. The TESS dataset (red models) is used to refine the ephemeris, with a floating dilution term, which is constrained using the ground-based photometry (blue models).}. \label{fig:transits}
\end{figure}

\subsection{Ground-based photometric observations}\label{sec:photometry}
\subsubsection{0.6 m RBO}
We observed three transits of TOI-7149~b with the 0.6 m telescope at the Red Buttes Observatory (RBO) in Wyoming, USA \citep{kasper_remote_2016} on 2024 August 2, 2025 February 22 and 2025 March 10. We used the Bessell I filter with no on-chip binning and exposure times of 240 s for all observations. The plate scale for RBO is 0.56$\unit{\arcsec/pixel}$, with a science aperture radius of 4 -- 5 pixels (2.2\arcsec -- 2.8\arcsec) across the three transits. The lightcurves are shown in \autoref{fig:transits}.

\subsubsection{0.3 m TMMT}
We observed a transit on 2024 August 26 using the using the Three-hundred MilliMeter ($300 \unit{mm}$) Telescope \citep[TMMT;][]{monson_standard_2017} at  Las Campanas Observatory in Chile. We used the Bessell I filter, without any on-chip binning, and used an exposure time of 180 s.  The plate scale during these observations was $0.97 \unit{\arcsec/pixel}$, and the science aperture radius was 4.8\arcsec. The lightcurve is shown in \autoref{fig:transits}.

\subsubsection{0.3 m LCRO}
We also used the 0.3 m Las Campanas Remote Observatory (LCRO) telescope at the Las Campanas Observatory in Chile alongside TMMT on 2024 August 26. The data was taken unbinned, with a plate scale of 0.64\arcsec{} per pixel in SDSS \textit{i'} band, and reduced with a science aperture radius of 2.9\arcsec. The lightcurve is shown in \autoref{fig:transits}.

Data for RBO, TMMT and LCRO were reduced with aperture photometry using the custom \texttt{python} pipeline described in \cite{kanodia_searching_2024-1}.

\subsubsection{Diffuser-assisted 5.1 m Palomar WIRC}
We observed a transit of TOI-7149~b in J-band ($\lambda_{\textrm{cent}} = 1.25~\mu m$) with the Wide Field Infrared Camera \citep[WIRC;][]{2003SPIE.4841..451W} on the  5.1 m Hale telescope at Palomar Observatory in California, USA on 2025 May 18. The observations included an engineered diffuser to defocus the PSF ($\sim$ 3\arcsec{} FWHM), which allowed for longer exposure times and increased photometric stability \citep{stefansson_toward_2017}. The data cadence was 37 seconds, with 30 second exposures and $\sim$ 6 -- 7 seconds for readout and processing. It was taken unbinned with a plate scale of 0.249\arcsec/pixel, and was reduced using \texttt{AstroImageJ} \citep{collins_astroimagej_2017} with a science aperture radius of 19 pixels. The lightcurves are shown in \autoref{fig:transits}.

\subsection{High Contrast Imaging}\label{sec:speckle}

\begin{figure}[] 
\centering
\includegraphics[width=0.45\textwidth]{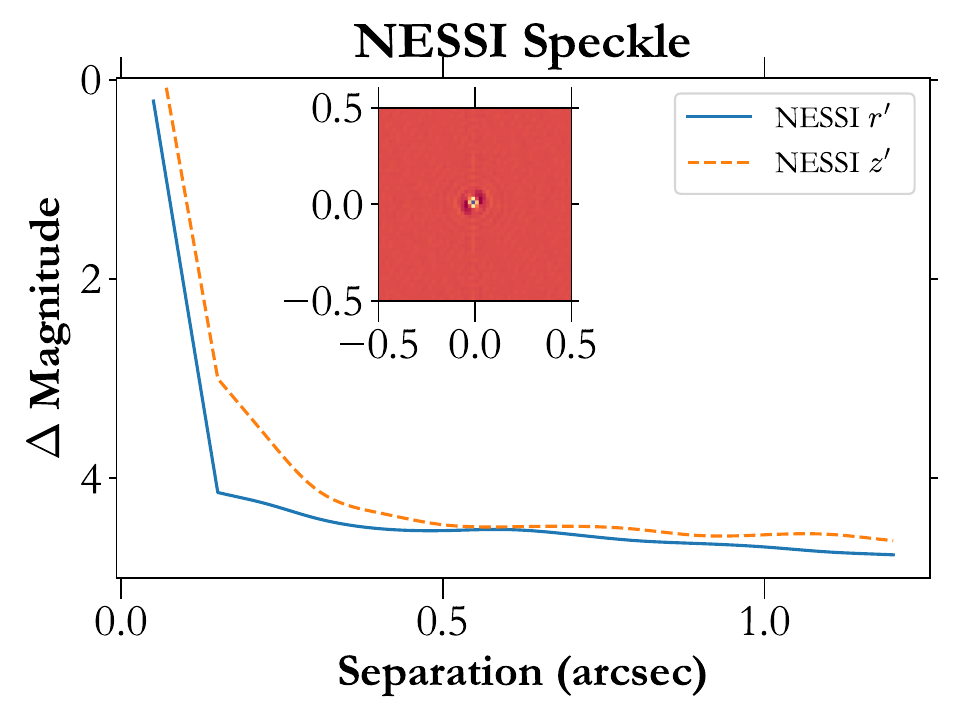}
\caption{5$\sigma$ contrast curve for TOI-7149 observed from NESSI in the Sloan \(r^\prime\)filter and \(z^\prime\)filter showing no bright companions within \(1.2''\) from the host star. The \(z^\prime\) image is shown as an inset 1$\arcsec$ across.} \label{fig:NESSI}
\end{figure}

We observed TOI-7149 using the NN-Explore Exoplanet Stellar Speckle Imager \citep[NESSI;][]{scott_nn-explore_2018} on the WIYN 3.5m telescope located at Kitt Peak National Observatory on the nights of 2024 September 10, 2025 February 18, and 2025 February 19 (2024B-103024, 2025A-103024: PI Kanodia). Sequences of 40 ms diffraction-limited frames were collected in the SDSS r' and z' filters using the blue and red NESSI cameras, respectively. The individual frames were combined to produce reconstructed speckle images following the methods of \citet{howell_speckle_2011}.
We determined that image quality of the 2025 February 18 observation was higher than on the other nights, which were affected by poor observing conditions, so we elect to use this night for the remainder of our analysis.
The speckle images and $5\sigma$ contrast limits are shown in Figure \ref{fig:NESSI}. We are able to exclude companions and background sources down to contrast limits of $\Delta m_{r'}=4.4$ and $\Delta m_{z'}=4.1$ at 0.3\arcsec and $\Delta m_{r'}=4.7$ and $\Delta m_{z'}=4.6$ at 1.0\arcsec, confirming that the ground-based observations do not have additional sources of dilution.

\subsection{Near-infrared spectroscopy with HPF}\label{sec:hpfrvs}

\begin{figure*}[!t]
\begin{center}
\fig{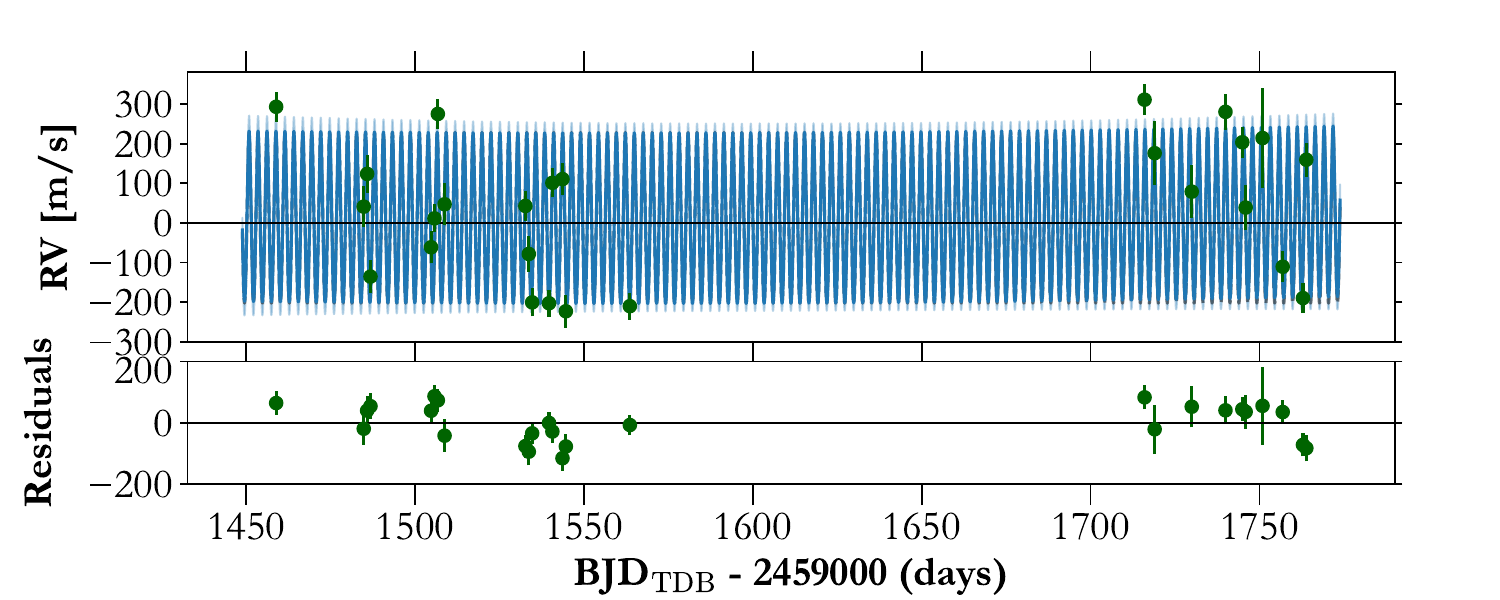}{0.63\textwidth}{}
\fig{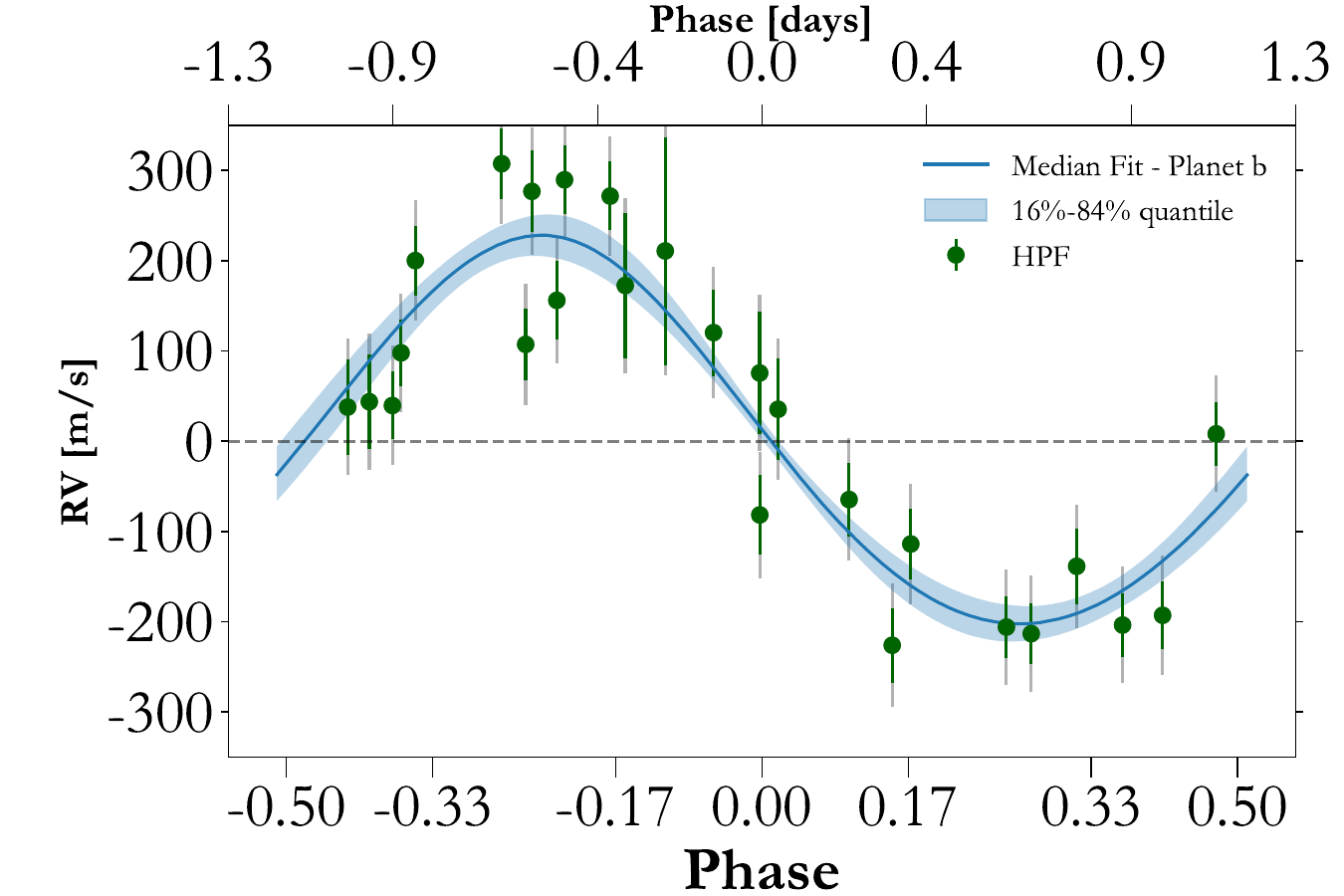}{0.36\textwidth}{}
\end{center}
% \vspace{-2 cm}
\caption{\small \textbf{Left:} Time series RVs of TOI-7149 with HPF data shown in green, and the best-fitting model derived from the joint fit to the photometry and RVs plotted in blue. The model includes the 16-84$\%$ confidence interval in lighter blue, as well as the RV trend detailed in \autoref{tab:planetprop}. The bottom panel shows the residuals after subtracting the model from the data. \textbf{Right:} HPF RV observations phase-folded to the best fit orbital period from the joint fit described in Section \ref{sec:joint}. Instrument errorbars are shown in green, while RV jitter added in quadrature is shown in the background on each point in grey.}\label{fig:hpf_rvs}
\end{figure*}

We obtained 25 visits on TOI-7149~b using HPF \citep[][]{mahadevan_habitable-zone_2012, mahadevan_habitable-zone_2014} between 2024 May 28 and 2025 March 29, which are shown in \autoref{fig:hpf_rvs}. HPF is a near-infrared (\(8080-12780\)\ \AA) spectrograph on the 10 m Hobby-Eberly Telescope\footnote{Based on observations obtained with the Hobby-Eberly Telescope (HET), which is a joint project of the University of Texas at Austin, the Pennsylvania State University, Ludwig-Maximillians-Universitaet Muenchen, and Georg-August Universitaet Goettingen. The HET is named in honor of its principal benefactors, William P. Hobby and Robert E. Eberly.} \citep[HET;][]{ramsey_early_1998, hill_hetdex_2021} in West Texas, USA. It is environmentally stabilized \citep{stefansson_versatile_2016} and has fiber-fed illumination \citep{kanodia_overview_2018, kanodia_harsh_2021}.  We corrected for bias, non-linearity, cosmic rays and processed the HPF slope images using the \texttt{HxRGproc} package \citep{ninan_habitable-zone_2018}. Given the faintness of the star and to avoid stray light contamination, we do not use the simultaneous calibration using the near-infrared Laser Frequency comb \citep[LFC;][]{metcalf_stellar_2019}. Instead, we interpolated the wavelength solution from other LFC exposures on the night of the observation, which has been shown to provide precise wavelength calibration and drift correction at the $\sim$ 30 \cms{} level \citep{stefansson_sub-neptune-sized_2020}. 

\begin{figure}[!t] 
\centering
\includegraphics[width=0.45\textwidth]{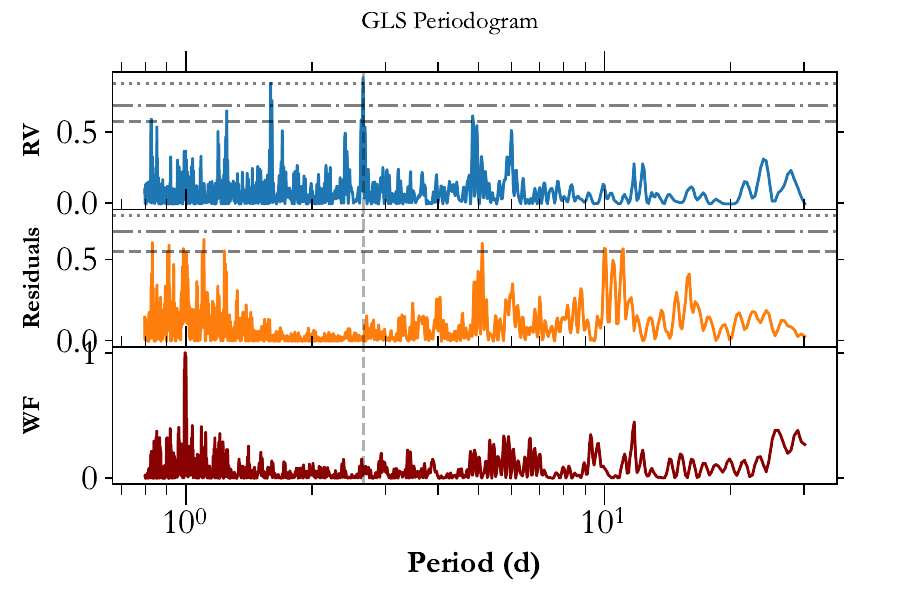}
\caption{\textbf{Top:} GLS of the HPF RVs in blue, with the dashed vertical line showing the planetary orbital period. The horizontal lines are the 10\%, 1\%, and 0.1\% False Alarm Probability levels. There is a peak in the RV periodogram, corresponding to the orbital period, that is significant at the $< 0.1\%$ level. \textbf{Middle:} GLS of the residuals obtained after subtracting the best-fit model do not display any significant peaks. \textbf{Bottom:} Window Function of the HPF RVs showing the characteristic 1.0 day peak indicative of observing cadence at HET.} \label{fig:rv_gls}
\end{figure}

We use the template-matching method \citep{anglada-escude_harps-terra_2012} implemented under the \texttt{SpEctrum Radial Velocity AnaLyser} pipeline \citep[\texttt{SERVAL};][]{zechmeister_spectrum_2018, stefansson_neptune-mass_2023} to extract RVs, which has been modified for HPF \citep{stefansson_sub-neptune-sized_2020}.  We used \texttt{barycorrpy} \citep{kanodia_python_2018} to perform the barycentric correction on the individual spectra, which is the Python implementation  of the algorithms from \cite{wright_barycentric_2014}.  Each visit consisted of two exposures of 969 seconds each that were subsequently combined by weighted averaging. The median S/N at 1070 nm per exposure is 21 per 1D extracted pixel, and the median RV uncertainty per visit (binned) is $\sim$ 40 \ms{}. The RVs used in our analysis are listed in \autoref{tab:rvs}. A generalised Lomb Scargle (GLS) periodogram \citep{lomb_least-squares_1976,scargle_studies_1982, zechmeister_generalised_2009} of the RVs (\autoref{fig:rv_gls}) shows a significant peak at the planetary orbital period ($<$ 0.1\% False Alarm Probability).

\begin{deluxetable}{ccc}
\tablecaption{RVs (binned in $\sim$ 30 minute exposures) of TOI-7149. \label{tab:rvs}}
\tablehead{\colhead{$\unit{BJD_{TDB}}$}  &  \colhead{RV}   & \colhead{$\sigma$}  \\
           \colhead{(d)}   &  \colhead{\ms{}} & \colhead{\ms{}}}
\startdata
2460458.89432 & 244.11 & 38.14 \\
2460484.82215 & -7.78 & 52.54 \\
2460485.82212 & 74.79 & 47.94 \\
2460486.81456 & -184.16 & 41.68 \\
2460504.75701 & -110.25 & 40.53 \\
2460505.76022 & -37.42 & 35.21 \\
2460506.75608 & 226.04 & 38.05 \\
2460508.74988 & -1.80 & 52.77 \\
2460532.68146 & -6.02 & 37.62 \\
2460533.68560 & -127.25 & 44.42 \\
2460534.67720 & -249.30 & 35.16 \\
2460539.66378 & -251.55 & 34.12 \\
2460540.66150 & 52.33 & 37.00 \\
2460543.65399 & 61.72 & 40.08 \\
2460544.65574 & -271.69 & 41.43 \\
2460563.60016 & -258.84 & 34.07 \\
2460715.97255 & 261.99 & 39.13 \\
2460718.96233 & 127.00 & 80.73 \\
2460729.93788 & 30.04 & 67.57 \\
2460739.92359 & 231.29 & 45.14 \\
2460744.90921 & 154.57 & 38.72 \\
2460745.90062 & -10.17 & 56.00 \\
2460750.89637 & 165.26 & 126.27 \\
2460756.87158 & -159.44 & 39.20 \\
2460762.86392 & -238.50 & 37.73 \\
2460763.86039 & 110.51 & 43.37 \\
\enddata
\end{deluxetable}

\begin{deluxetable*}{lccc}
\tablecaption{Summary of stellar parameters for TOI-7149. \label{tab:stellarparam}}
\tablehead{\colhead{~~~Parameter}&  \colhead{Description}&
\colhead{Value}&
\colhead{Reference}}
\startdata
\multicolumn{4}{l}{\hspace{-0.2cm} Main identifiers:}  \\
~~~TOI & \tess{} Object of Interest & 7149 & \tess{} mission \\
~~~TIC & \tess{} Input Catalogue  & 459323923 & Stassun \\
~~~2MASS & \(\cdots\) & J16185724+1910327 & 2MASS  \\
~~~Gaia DR3 & \(\cdots\) & 1200751810900182400 & Gaia DR3\\
\multicolumn{4}{l}{\hspace{-0.2cm} Equatorial Coordinates and Proper Motion:} \\
~~~$\alpha_{\mathrm{J2000}}$ &  Right Ascension (RA) & 244.738 $\pm$ 0.034 & Gaia DR3\\
~~~$\delta_{\mathrm{J2000}}$ &  Declination (Dec) & 19.176 $\pm$ 0.028 & Gaia DR3\\
~~~$\mu_{\alpha}$ &  Proper motion (RA, \unit{mas/yr}) &  -7.698 $\pm$ 0.039 & Gaia DR3\\
~~~$\mu_{\delta}$ &  Proper motion (Dec, \unit{mas/yr}) & 12.882 $\pm$ 0.041 & Gaia DR3 \\
~~~$\varpi$ &  Parallax in mas  & $7.86 \pm 0.05$  & Gaia DR3 \\
~~~$G$ & $G$ mag & $16.1324\pm0.0007$ & Gaia DR3 \\
~~~\(A_{V,\mathrm{max}}\) & Maximum visual extinction & $0.274$ & Green\\
\multicolumn{4}{l}{\hspace{-0.2cm} Optical and near-infrared magnitudes:}  \\
~~~$g^{\prime}$ &  PS1 $g^{\prime}$ mag  & $18.08 \pm 0.01$ & PS1\\
~~~$r^{\prime}$ &  PS1 $r^{\prime}$ mag  & $16.87 \pm 0.02$ & PS1 \\
~~~$i^{\prime}$ &  PS1 $i^{\prime}$ mag  & $15.409 \pm 0.005$ & PS1 \\
~~~$z^{\prime}$ &  PS1 $z^{\prime}$ mag  & $14.746 \pm 0.008$ & PS1 \\
~~~$y^{\prime}$ &  PS1 $y^{\prime}$ mag  & $14.41 \pm 0.01$ & PS1\\
~~~$J$ & $J$ mag & $13.16 \pm 0.03$ & 2MASS\\
~~~$H$ & $H$ mag & $12.47 \pm 0.03$ & 2MASS\\
~~~$K_s$ & $K_s$ mag & $12.23 \pm 0.02$ & 2MASS\\
~~~$W1$ &  WISE1 mag & $12.027 \pm 0.098$ & WISE\\
~~~$W2$ &  WISE2 mag & $11.914 \pm 0.022$ & WISE\\
~~~$W3$ &  WISE3 mag & $12.045 \pm 0.284$ & WISE\\
\multicolumn{4}{l}{\hspace{-0.2cm} Spectroscopic parameters$^a$:}\\
~~~$T_{\mathrm{eff}}$ &  Effective temperature in \unit{K} & $3363\pm59$ & This work\\
~~~$\mathrm{[Fe/H]}$ &  Metallicity in dex & $-0.08\pm0.16$ & This work\\
~~~$\log(g)$ & Surface gravity in cgs units & $4.88\pm0.04$ & This work\\
~~~$v\sin i_{\star}$ & Rotational broadening (km s$^{-1}$) & $<2$ & This work\\
\multicolumn{4}{l}{\hspace{-0.2cm} Model-dependent stellar SED and isochrone fit parameters$^b$:}\\
~~~$M_*$ &  Mass in $M_{\odot}$ & $0.344\pm0.030$ & This work\\
~~~$R_*$ &  Radius in $R_{\odot}$ & $0.351\pm0.015$ & This work\\
~~~$L_*$ &  Luminosity in $L_{\odot}$ & $0.01215\pm0.00063$ & This work\\
~~~$\rho_*$ &  Density in $\unit{g/cm^{3}}$ & $11.2\pm1.0$ & This work\\
~~~Age & Age in Gyrs & $7.8\pm5.0$ & This work\\
~~~Distance & Distance in pc & $127.24\pm0.74$ & This work\\
~~~$A_v$ & Visual extinction in mag & $0.100^{+0.037}_{-0.058}$ & This work\\
\multicolumn{4}{l}{\hspace{-0.2cm} Other Stellar Parameters:}           \\
~~~$\Delta$RV &  ``Absolute'' radial velocity in \unit{km/s} & $-36.7\pm0.5$ & This work\\
~~~$U, V, W$ &  Galactic velocities in \unit{km/s} &  $-26.47\pm0.31, -7.67\pm0.22, -25.87\pm0.33$ & This work\\
~~~$U, V, W^c$ &  Galactic velocities (LSR) in \kms{} & $-15.37\pm0.90, 4.57\pm0.72, -18.62\pm0.69$ & This work\\
\enddata
\tablenotetext{}{References are: Gaia DR3 \citep{gaia_collaboration_gaia_2023}, Green \citep{green_3d_2019}, Stassun \citep{stassun_tess_2018}, 2MASS \citep{cutri_2mass_2003}, PS1 \citep[DR2;][]{chambers_pan-starrs1_2016} }
\tablenotetext{a}{Derived using the HPF spectral matching algorithm from \cite{stefansson_sub-neptune-sized_2020}}
\tablenotetext{b}{{\tt EXOFASTv2} derived values using MIST isochrones with the \gaia{} parallax and spectroscopic parameters in $a$) as priors.}
\tablenotetext{c}{The barycentric UVW velocities are converted into local standard of rest (LSR) velocities using the constants from \cite{schonrich_local_2010}.}

\end{deluxetable*}

\section{Stellar Parameters}\label{sec:stellar}

\subsection{Spectroscopic parameters from HPF}
We used the \texttt{HPF-SpecMatch}\footnote{\url{https://gummiks.github.io/hpfspecmatch/}} package \citep[][]{Stefansson2020} to derive the stellar effective temperature ($T_e$), surface gravity ($\log g_\star$), metallicity ([Fe/H]), and rotational velocity ($v\sin i_\star$) from the observed high-resolution spectra of TOI-7149. \texttt{HPF-SpecMatch} empirically derives spectroscopic parameters \citep[e.g.,][]{Yee2017} by using a weighted linear combination of the five stellar spectra from a library of well-characterized stars that best-match the target star. The HPF spectral library used in this work consisted of 100 stars with parameters covering $2700\mathrm{K} \le T_{e} \le 4500~\mathrm{K}$, $4.63<\log g_\star < 5.26$, and $-0.49 < \mathrm{[Fe/H]} < 0.53$. We analyzed the spectral region between $8534-8645$ \AA{} because of the minimal telluric contamination. The uncertainties reported were the standard deviation of the residuals from a leave-one-out cross-validation procedure applied to the library \citep[see more details in][]{Stefansson2020}. Our analysis determined $T_{e}=3363\pm59$ K, $\log g_\star=4.88\pm0.04$, and $\mathrm{[Fe/H]}=-0.08\pm0.16$ for TOI-7149. \texttt{HPF-SpecMatch} could only place a upper limit of $v\sin i_\star<2~\mathrm{km~s^{-1}}$, which is consistent with the lack of rotation signal in TESS or HPF activity indicators, thereby suggesting an old inactive star ($>$ 1 Gyr).

\subsection{Stellar parameters from the spectral energy distribution}

We followed the procedures in \cite{canas_toi-3984_2023} to derive stellar parameters using the \texttt{EXOFASTv2} analysis package \citep{Eastman2019} to model the spectral energy distribution (SED). \texttt{EXOFASTv2} models the observed magnitudes using the MIST model grids \citep{dotter_mesa_2016,choi_mesa_2016} and uses the $R_{v}=3.1$ reddening law from \cite{fitzpatrick_correcting_1999} to calculate a visual magnitude extinction. We applied Gaussian priors on the (i) broadband optical and near-infrared photometry listed in \autoref{tab:stellarparam}, (ii) spectroscopic parameters from \texttt{HPF-SpecMatch}, and (iii) parallax from Gaia DR3 \citep{gaia_collaboration_gaia_2023}. We used a uniform prior on the visual extinction ($A_V$) with an upper limit derived from the estimates by \cite{green_3d_2019} at the location of TOI-7149. The stellar parameters are presented in  \autoref{tab:stellarparam}. TOI-7149 has a mass and radius of $M_\star=0.344\pm0.030~\mathrm{M_\odot}$ and $R_\star=0.351\pm0.015~\mathrm{R_\odot}$, respectively. The \teff{} estimate from SED photometry is 3237 $\pm$ K, which is about 2-$\sigma$ lower than our spectroscopic estimate. However, for this work we choose the spectroscopic \texttt{HPF-SpecMatch} estimates.

We use the colour-magnitude relations in Table 4 from \cite{kiman_exploring_2019} and Table 7 from \cite{cifuentes_carmenes_2020} to estimate the spectral subtype of TOI-7149 using Gaia and 2MASS photometry as a fully convective M4 dwarf.

\subsection{Galactic kinematics}

We use the systemic velocity from HPF and astrometry from Gaia DR3 to calculate the \textit{UVW} velocities\footnote{With \textit{U} towards the Galactic center, \textit{V} towards the direction of Galactic spin, and \textit{W} towards the North Galactic Pole  \citep{johnson_calculating_1987}.} in the barycentric frame using \texttt{GALPY} \citep{bovy_galpy_2015}, which are listed in \autoref{tab:stellarparam}. Using the BANYAN tool \citep{gagne_banyan_2018}, we classify TOI-7149 as a field star in the thin disk \citep[][]{bensby_exploring_2014}.

\subsection{Stellar companions}
Gaia DR3 \citep{gaia_collaboration_gaia_2023} provides an astrometric constraint on the lack of unresolved companions using the re-normalised unit weight error (RUWE) metric. For TOI-7149, the reported RUWE is 0.973, which is lower than the commonly accepted threshold in literature of $\gtrsim$ 1.4 to ascertain the potential presence of stellar companions in binary studies \citep{penoyre_binary_2020, belokurov_unresolved_2020}. Similarly, there is no significant astrometric excess noise reported alongside the Gaia astrometry.  Lastly, the host star is not part of a bound pair as determined from the \gaia{} catalogue of \cite{el-badry_million_2021}. This is consistent with the lack of any bright companions from speckle imaging in \S\ref{sec:speckle} and, together, suggests that TOI-7149 is a single-star.

\section{Joint Fitting of Photometry and RVs}\label{sec:joint}
Similar to previous discoveries from our \textit{Searching for GEMS} survey, we perform a joint fit of the photometry and RVs using the \texttt{python} package \texttt{exoplanet} \citep{foreman-mackey_exoplanet-devexoplanet_2021} which uses \texttt{PyMC3}, the Hamiltonian Monte Carlo (HMC) package \citep{salvatier_probabilistic_2016}. We use a quadratic limb-darkening law with coefficients derived using the reparametrization suggested by \cite{kipping_efficient_2013}. Each instrument (TESS, Palomar, RBO, LCRO, TMMT) has separate limb-darkening coefficients. Due to TESS' large pixels and the presence of background stars, we included a dilution term for each TESS sector and placed a uniform prior between 0.001 and 1.5 \citep[explained in][]{kanodia_toi-5205b_2023}, such that the transit depth was determined using the other photometric datasets where the stars are spatially well-resolved and excluded from the apertures used to derive the photometry. We also included an RV jitter (white noise), RV offset ($\gamma$) and quadratic RV trend in the fit. The orbital eccentricity is consistent with circular ($<$ 0.12, $<$ 0.17, $<$ 0.22, at 1-, 2-, 3-$\sigma$, respectively), especially when considering the Lucy-Sweeney bias \citep{lucy_spectroscopic_1971}. The derived parameters for this system are listed in \autoref{tab:planetprop}. The measured RV trend is consistent with 0 (i.e., a flat line, at $<$ 1 $\sigma$). The phase-folded transit photometry and best fit models (and uncertainty) are shown in \autoref{fig:transits}. The RV time series and phase-folded fit are shown in \autoref{fig:hpf_rvs}.

% \startlongtable
\begin{deluxetable*}{llc}
\tablecaption{Derived Parameters for the TOI-7149 System.   \label{tab:planetprop}}
\tablehead{\colhead{~~~Parameter} &
\colhead{Units} &
\colhead{Value$^a$} 
}
\startdata
\sidehead{Orbital Parameters:}
~~~Orbital Period\dotfill & $P$ (days) \dotfill & 2.65206166 $\pm 0.00000080$\\
~~~Eccentricity\dotfill & $e$ \dotfill & 0.078$^{+0.047}_{-0.048}$ \\
~~~Argument of Periastron\dotfill & $\omega$ (radians) \dotfill & -0.286$^{+0.692}_{-0.587}$ \\
~~~Semi-amplitude Velocity\dotfill & $K$ (\ms{})\dotfill &
216 $\pm$ 21\\
~~~Systemic Velocity$^b$\dotfill & $\gamma_{\mathrm{HPF}}$ (\ms{})\dotfill & -49 $\pm$ 17\\
~~~RV trend\dotfill & $dv/dt$ (\ms{} yr$^{-1}$)   & 13$^{+44}_{-42}$   \\ 
~~~ \dotfill & $d^2v/dt^2$ (\ms{} yr$^{-2}$)   & 52$^{+96}_{-98}$   \\ 
~~~RV jitter\dotfill & $\sigma_{\mathrm{HPF}}$ (\ms{})\dotfill & 55$^{+15}_{-13}$\\
\sidehead{Transit Parameters:}
~~~Transit Midpoint \dotfill & $T_C$ (BJD\textsubscript{TDB})\dotfill & 2459703.596050 $\pm$ 0.00029\\
~~~Scaled Radius\dotfill & $R_{p}/R_{*}$ \dotfill & 0.3334$^{+0.0048}_{-0.0053}$\\
~~~Scaled Semi-major Axis\dotfill & $a/R_{*}$ \dotfill & 15.52$^{+0.67}_{-0.54}$\\
~~~Orbital Inclination\dotfill & $i$ (degrees)\dotfill & 89.25$^{+0.39}_{-0.28}$\\
~~~Transit Duration\dotfill & $T_{14}$ (days)\dotfill & 0.0719$^{+0.0025}_{-0.0030}$\\
~~~Dilution$^{c}$ \dotfill & $D_{\mathrm{TESS~S25}}$ \dotfill & $1.074\pm0.035$\\
~~~ & $D_{\mathrm{TESS~S51}}$ \dotfill & $1.17\pm0.04$\\ 
~~~ & $D_{\mathrm{TESS~S52}}$ \dotfill & $1.21\pm0.04$\\ 
~~~ & $D_{\mathrm{TESS~S79}}$ \dotfill & $1.12\pm0.034$\\ 
~~~Limb Darkening$^d$ $\dotfill$ & $(u_{1}$, $u_{2}$)$_{\mathrm{TESS}}$ $\dotfill$ & 0.36$^{+0.27}_{-0.23}$, $0.47^{+0.28}_{-0.38}$  \\
~~~ $\dotfill$ & $(u_{1}$, $u_{2}$)$_{\mathrm{RBO}}$ $\dotfill$ & 0.19$^{+0.19}_{-0.13}$, $0.26^{+0.29}_{-0.30}$  \\
~~~ $\dotfill$ & $(u_{1}$, $u_{2}$)$_{\mathrm{TMMT}}$ $\dotfill$ & 0.37$^{+0.33}_{-0.26}$, $0.10^{+0.37}_{-0.30}$  \\
~~~ $\dotfill$ & $(u_{1}$, $u_{2}$)$_{\mathrm{LCRO}}$ $\dotfill$ & 0.37$^{+0.32}_{-0.25}$, $0.06^{+0.35}_{-0.26}$ \\
~~~ $\dotfill$ & $(u_{1}$, $u_{2}$)$_{\mathrm{Palomar}}$ $\dotfill$ & $0.28 \pm 0.17$, $0.19^{+0.32}_{-0.29}$  \\
\sidehead{Planetary Parameters:}
~~~Mass\dotfill & $M_{p}$ (M$_\oplus$)\dotfill &  $224\pm24$\\
~~~ & $M_{p}$ ($M_J$)\dotfill &  $0.705\pm0.075$\\
~~~Radius\dotfill & $R_{p}$  (R$_\oplus$) \dotfill& $13.2 \pm 0.51$\\
~~~ & $R_{p}$  ($R_J$) \dotfill& 1.18 $\pm$ 0.045\\
~~~Density\dotfill & $\rho_{p}$ (\gcmcubed{})\dotfill & 0.530$^{+0.085}_{-0.069}$\\
~~~Semi-major Axis\dotfill & $a$ (AU) \dotfill & 0.02603$^{+0.00071}_{-0.00076}$\\
~~~Average Incident Flux$^e$\dotfill & $\langle F \rangle$ (\unit{10^5\ W/m^2})\dotfill &  0.280 $\pm$ 0.040\\
~~~Planetary Insolation & $S$ (S$_\oplus$)\dotfill &  20.6 $\pm$ 2.9\\
~~~Equilibrium Temperature$^f$ \dotfill & $T_{\mathrm{eq}}$ (K)\dotfill & 593 $\pm$ 21\\
\enddata
\tablenotetext{a}{The reported values refer to the 16-50-84\% percentile of the posteriors.}
\tablenotetext{b}{RV offset in addition to the ``Absolute RV" from \autoref{tab:stellarparam}.}
\tablenotetext{c}{Dilution over-correction in \texttt{tglc} from background stars in \tess{}, constrained using ground-based apertures.}
\tablenotetext{d}{Where $u_1 + u_2 < 1$, and $u_1 > 0$ according to \cite{kipping_efficient_2013}.}
\tablenotetext{e}{We use a Solar flux constant = 1360.8 W/m$^2$, to convert insolation to incident flux.}
\tablenotetext{f}{We assume the planet to be a black body with zero albedo and perfect energy redistribution to estimate the equilibrium temperature. }
\normalsize
\end{deluxetable*}

\begin{figure*}[!t]
\fig{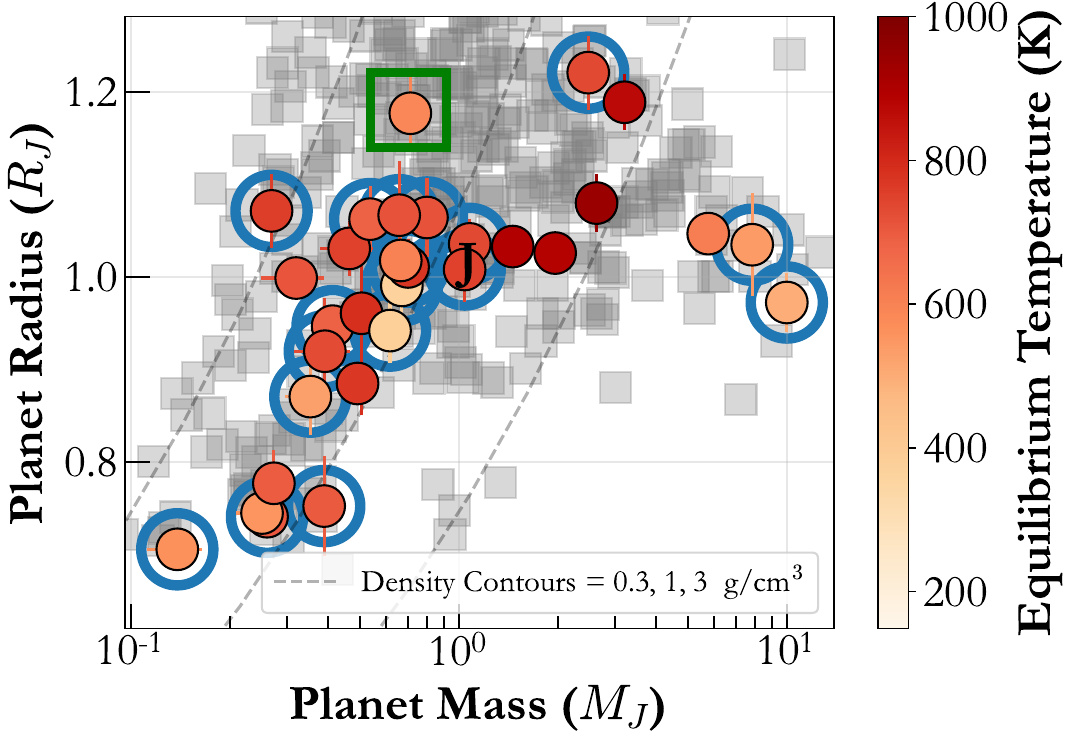}{0.47\textwidth}
{}    \label{fig:RadiusMass}
\fig{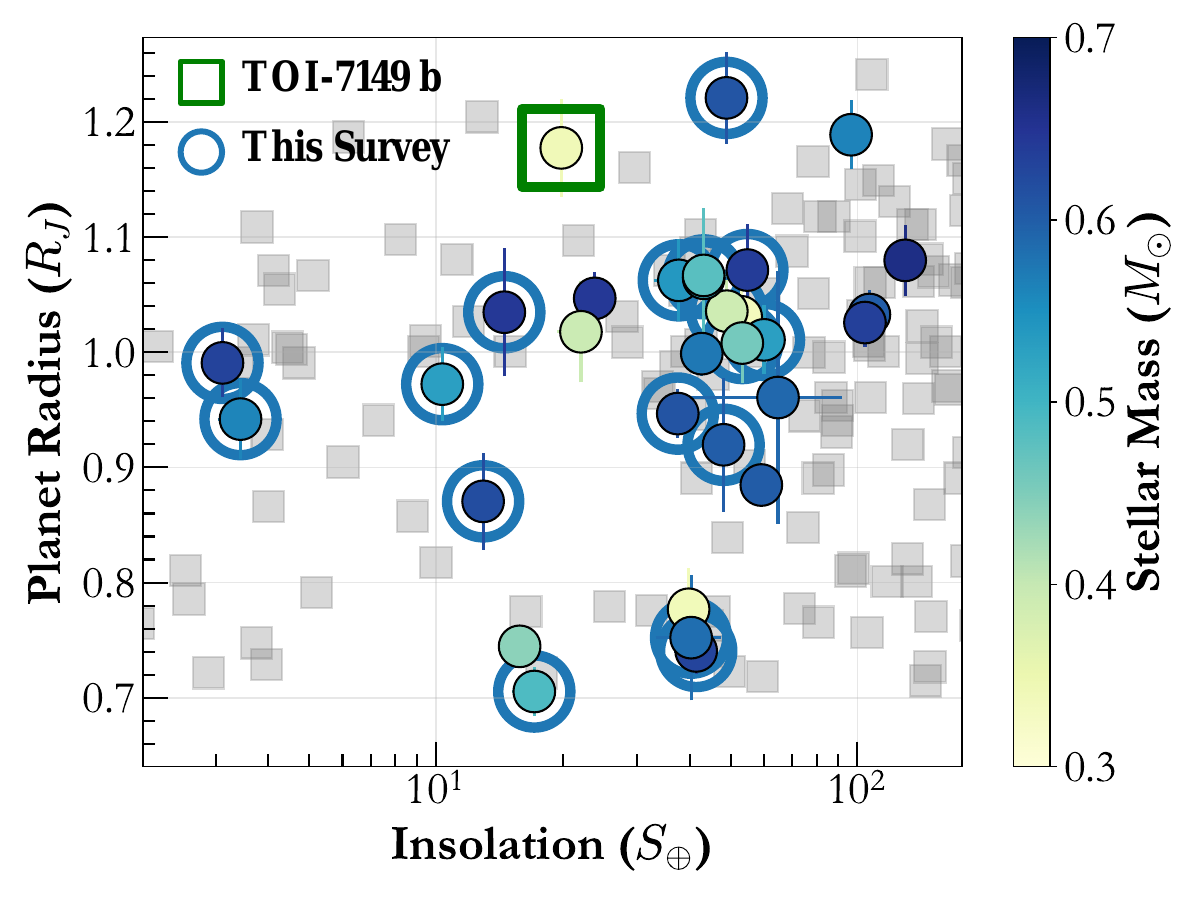}{0.45\textwidth}
{} \label{fig:RadiusPeriod} \vspace{0.2 cm}
\fig{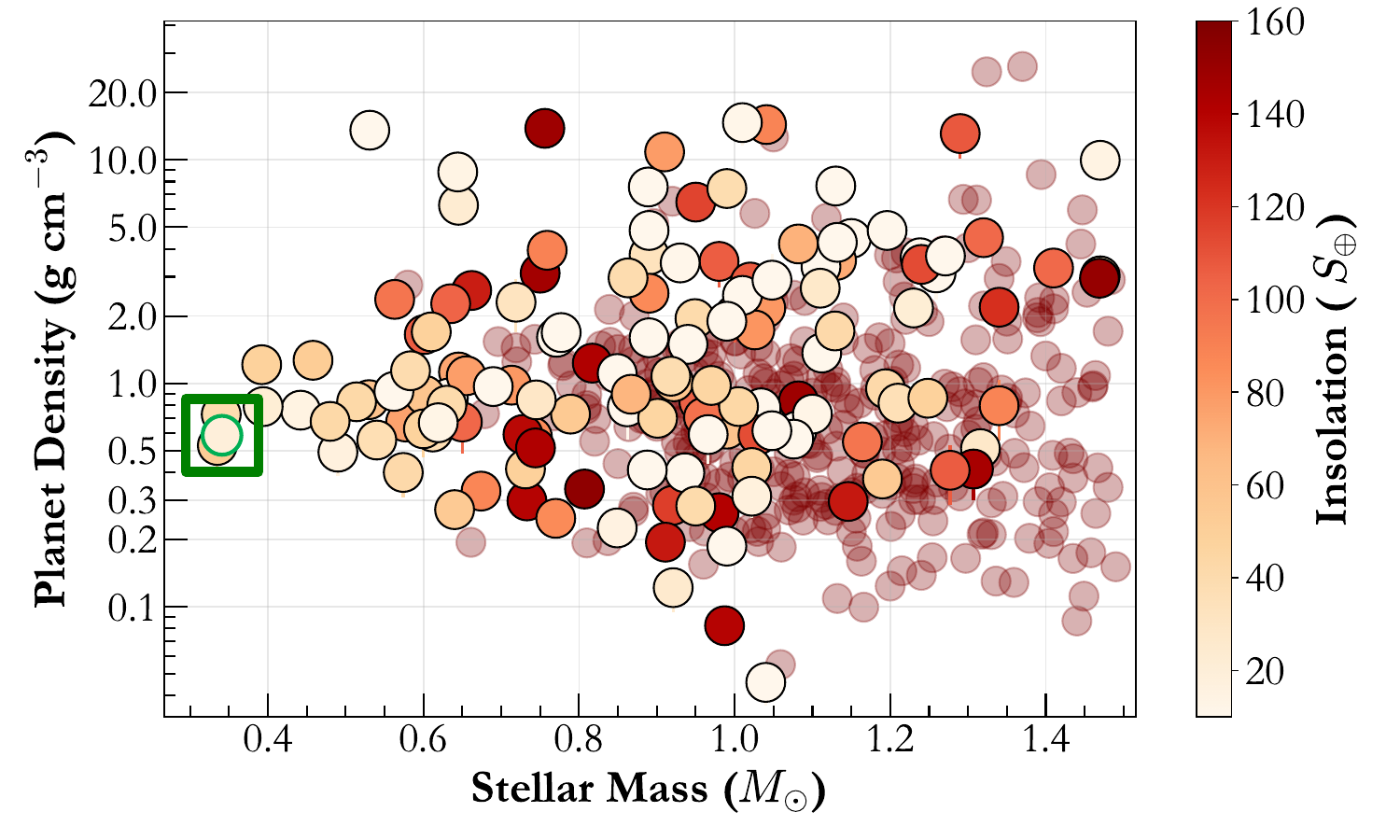}{0.90\textwidth \hspace{0.3 cm}}
{} \label{fig:InsolationStellarMass}
\caption{\small \textbf{Top Left:} TOI-7149~b (green square) in a mass-radius plane alongside other transiting M dwarf planets (coloured by equilibrium temperature). The GEMS discovered by the \textit{Searching for GEMS} survey are circled in blue. We also include planets around FGK stars in the background, along with density contours for 0.3, 1, 3 \gcmcubed{} \citep{PSCompPars}. \textbf{Top Right:} The insolation-radius plane is shown for the same sample of planets. \textbf{Bottom:} Bulk planetary density vs stellar mass, colour coded by insolation flux. Warm Jupiters orbiting FGK dwarfs receiving $< 160~S_{\oplus}$ (T$_{\rm{eq}}$ $\sim$ 1000 K) flux are shown in solid colour, with hot Jupiters in the background (red circles). TOI-7149 is one of the lowest mass stars hosting a transiting giant planet.}\label{fig:paramspace}
\end{figure*}

\section{Discussion}\label{sec:discussion}
\subsection{TOI-7149~b compared to other GEMS}

\begin{figure*}[!t]
\begin{center}
\hspace{-2cm}
\fig{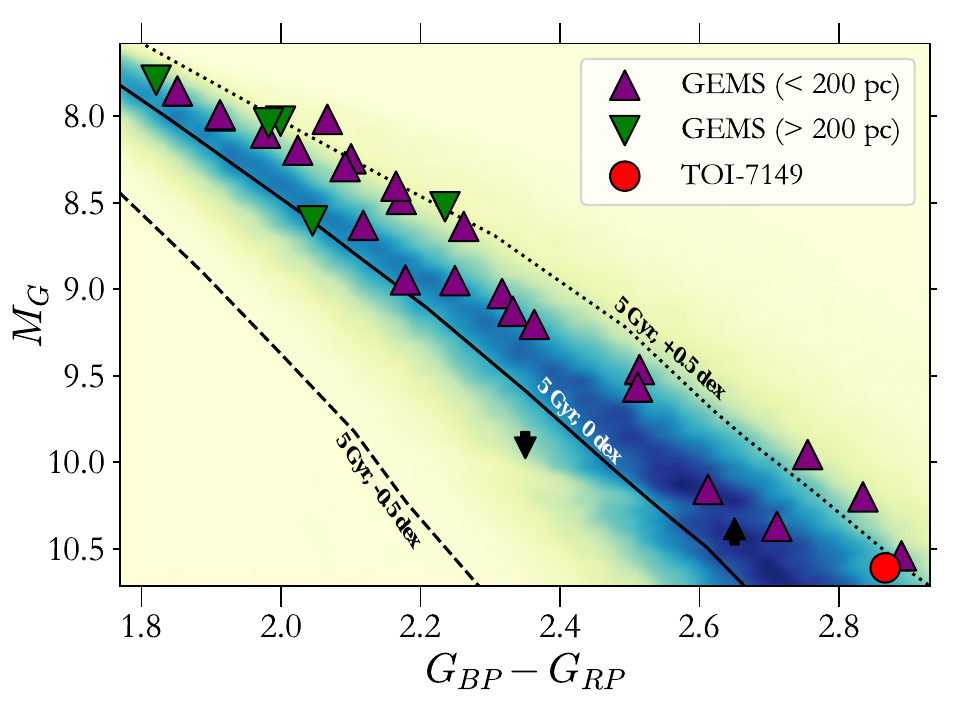}{0.50\textwidth}{}
\hspace{-4cm}
\fig{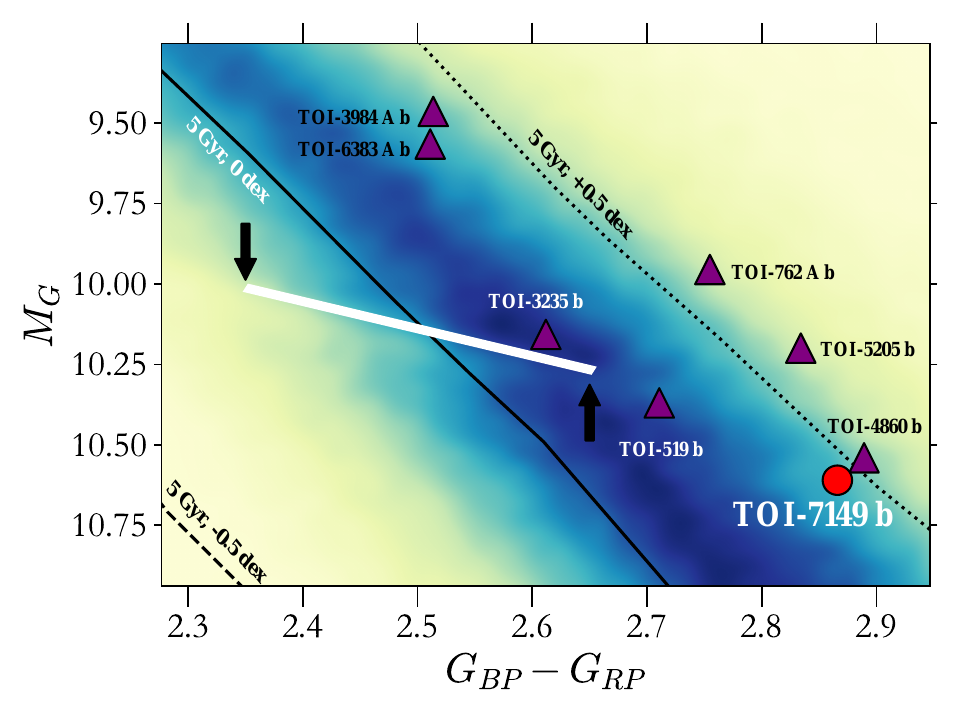}{0.50\textwidth}{}
\end{center}
\vspace{-1 cm}
\caption{\small We show a Gaia colour-magnitude diagram (CMD) for the 200 pc sample from the \textit{Searching for GEMS} survey with a Gaussian KDE, which is darker in regions with more stars. The upright violet triangles indicate GEMS within 200 pc, whereas the inverted green ones have distances $>$ 200 pc. The red dot shows TOI-7149, which is the one of the coldest and reddest M-dwarfs to host one of these GEMS. We also overplot the PARSEC isochrones for a 5 Gyr system at +0.5, 0 and -0.5 dex metallicity \citep{bressan_parsec_2012, chen_improving_2014}. The white strip and black arrows indicate the upper edge of the convective transition between partially and fully convective M-dwarfs \citep{jao_gap_2018}.  The full extent of the CMD is shown on the left, whereas on the right we zoom into the region near the transition zone to show other similar stars to TOI-7149. }\label{fig:CMD}
\end{figure*}

To contextualize TOI-7149~b with respect to other transiting planets, especially transiting GEMS, we query the NASA Exoplanet Archive Planetary Systems Data Table on 2025 April 11 \citep{akeson_nasa_2013, PSCompPars}. To this, we add TOI-7149~b (this work), TOI-6330~b and TOI-6303~b \citep{hotnisky_searching_2024}, TOI-5573~b \citep{fernandes_searching_2025}, TOI-5349~b (Sandoval et al. in prep.), TOI-5916~b (O'Brien, Wong et al. in prep.)\footnote{This work also includes the discovery of TOI-6158~b, however it is grazing and does not meet our radius precision cuts.}. To focus on FGKM host stars, we exclude planets with hosts $>$ 7200 K or $> 1.5$ \solmass. Furthermore, we limit the sample to planets with radii $\gtrsim$ 8 \earthradius{} within 1-$\sigma$ ($\gtrsim$ 0.7 $R_J$), radius precision $>$~5-$\sigma$, and mass precision $>$~3-$\sigma$. This results in 544 transiting giant planets that are shown in \autoref{fig:paramspace}, of which $\sim$ 30 are transiting GEMS. We show that TOI-7149 is one of the lowest mass stars to host a transiting giant planet.

TOI-7149~b is a fully-convective M-dwarf, as can be seen from its location (\autoref{fig:CMD}) below the transition between partially and fully convective M-dwarfs \citep{jao_gap_2018}. This gap marks the convective instability brought about by fluctuations in $^3$He fusion and episodic cycling therein \citep{van_saders_3he-driven_2012, macdonald_explanation_2018, baraffe_closer_2018, feiden_gaia_2021}. Furthermore, this gap is associated with luminosity and temperature variations for stars within the narrow strip, as well as marked differences between the flare rates and activity levels on both sides of it \citep{jao_mind_2023, jao_mind_2025-1}. Additionally, \autoref{fig:CMD} shows that a number of transiting GEMS are above the main-sequence and may be metal-rich as indicated by their location with respect to the PARSEC\footnote{\url{http://stev.oapd.inaf.it/cmd}} \citep{bressan_parsec_2012, chen_improving_2014} isochrones. This apparent preference for GEMS to orbit metal-rich stars is consistent with findings from \cite{gan_metallicity_2024} and \cite{rodriguez_martinez_comparison_2023}. We note that TOI-7149 appears to be much more metal-rich in the CMD as compared to our \texttt{HPF-SpecMatch} metallicity estimates. However, we add the usual caveats regarding M-dwarf metallicities, both from empirical  methods such as \texttt{HPF-SpecMatch}, as well as synthetic isochrones. These are due to the faintness of the star, the molecular lines erasing the continuum \citep{passegger_metallicities_2022}. Furthermore, errors due to interpolating synthetic atmospheres are particularly egregious for M-dwarfs \citep{wheeler_korg_2024}. Therefore robust metallicity dependence studies for these planets necessitate a homogeneous analysis of the host-star metallicities within a consistent framework, that is currently beyond the scope of this work.

\subsection{TOI-7149~b appears to be inflated}
\begin{figure}[!b] 
\centering
\includegraphics[width=0.45\textwidth]{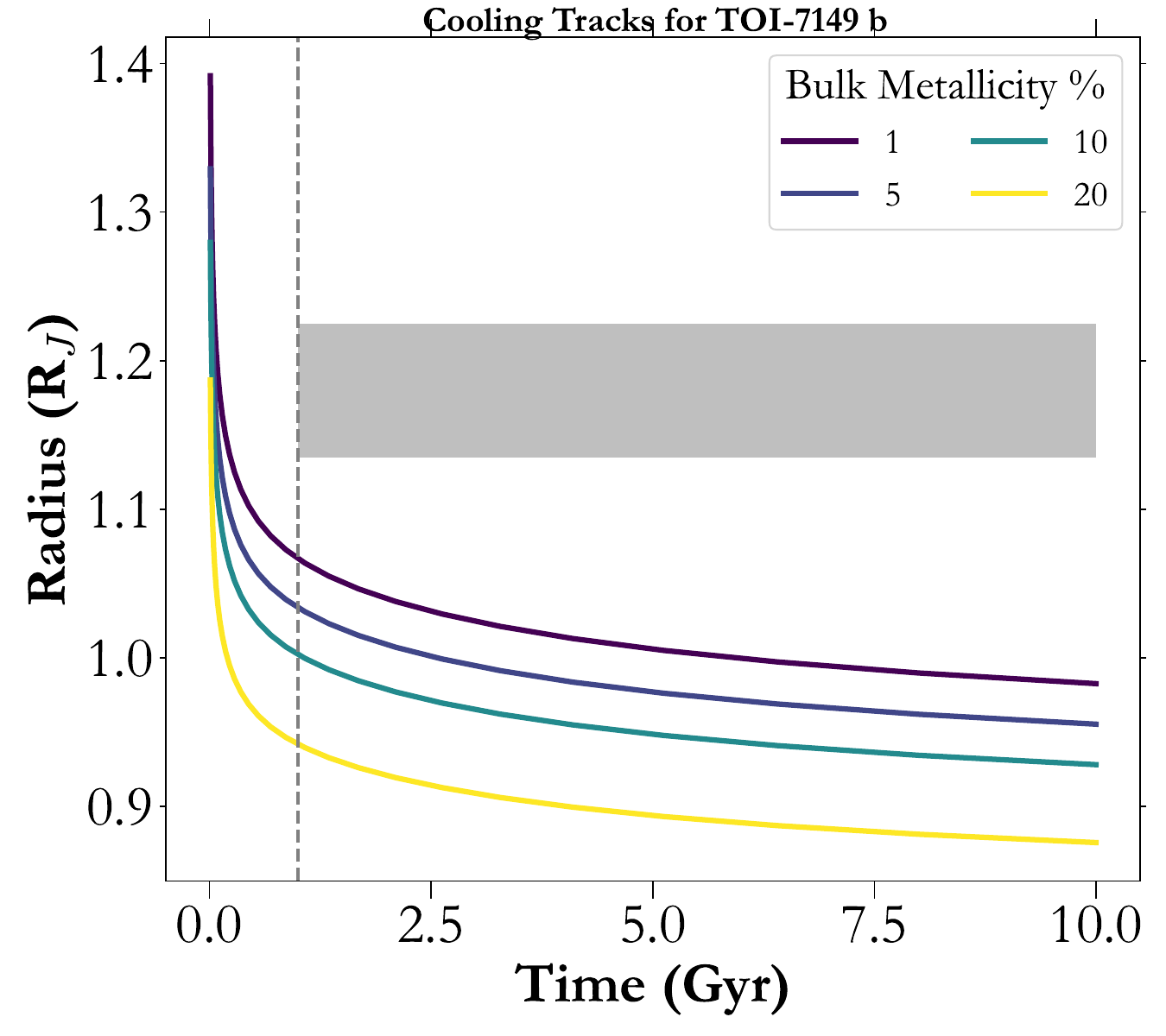}
\caption{The grey bar shows the 1-$\sigma$ radius measurement along with age estimates $>$ 1 Gyr. The different colour cooling tracks generated from \texttt{planetsynth} assume the minimum atmospheric metallicity of solar (1.2\%) and different bulk metallicities. \textbf{Takeaway:} The measured radius for TOI-7149~b is inconsistent with theoretical model predictions, thereby suggesting that it is inflated.} \label{fig:coolingtrack}
\end{figure}

TOI-7149~b appears to be quite inflated at $\sim$ 1.2 $R_J$, though the mechanism causing this for either planet is not currently known. We verified this by using \texttt{planetsynth} \citep{muller_synthetic_2021} synthetic planetary evolution tracks and predicting the radius for TOI-7149~b based on its mass, insolation flux, and assuming different bulk-metallicities (\autoref{fig:coolingtrack}). The predicted planetary radii ($\sim 1~R_J$) are all inconsistent with the measured radius of $1.18 \pm 0.045~R_J$ and the old age ($>>$~1~Gyr) of the host star suggested by its lack of activity. TOI-7149~b is hence likely inflated despite its insolation flux (20 $S_{\oplus})$ being much lower than the $160~S_{\oplus}$ threshold, above which radius inflation seen for hot Jupiters \citep{thorngren_bayesian_2018}. Thus it necessitates some additional source of heating, or delay in cooling to explain its structure.

\subsection{Future characterization to explore the inflated nature of TOI-7149~b}
\begin{figure}[!t] 
\centering
\includegraphics[width=0.5\textwidth]{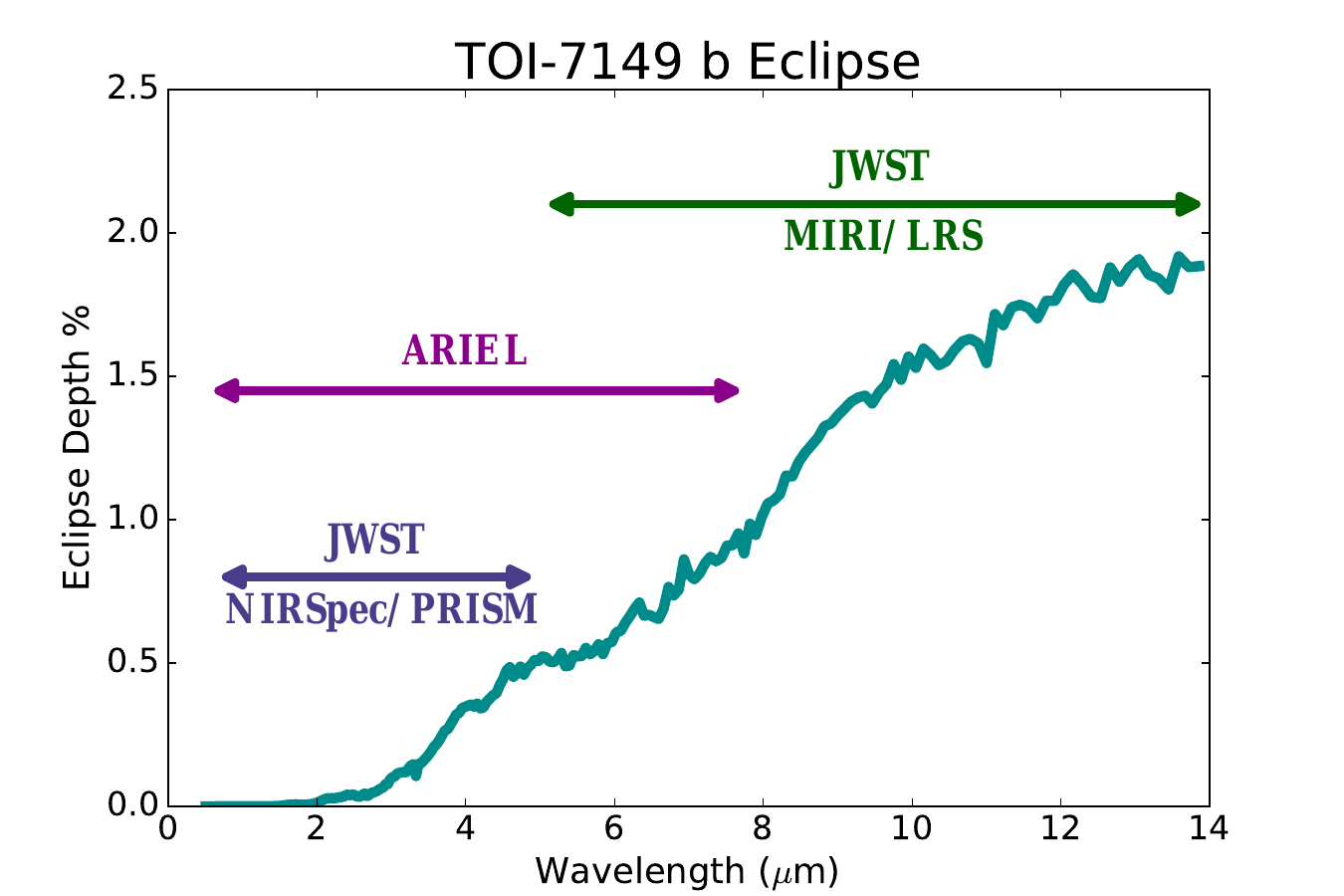}
\caption{Simulated emission spectrum of TOI-7149~b, also showing the instrument bandpasses for JWST NIRSpec and MIRI, along with ARIEL.} \label{fig:emission}
\end{figure}

The low density and inflated nature of TOI-7149~b make it a valuable target for future atmospheric characterization (TSM $\sim$ 132; ESM $\sim$ 78). Emission spectroscopy can measure the thermal emission from the planet and hence its effective temperature\footnote{Note that this is distinct from the stellar effective temperature, but refers to a similar concept for the planet as determined by its thermal emission. This \textit{can} be and is usually different from the equilibrium temperature ($T_{\rm{eq}}$), which in this context is akin to the irradiation temperature (or the expected temperature of the planet based on the flux it receives). If a planet has no internal source of heat ($T_{\rm{int}} \sim 0$), then \teff{} $\simeq$ $T_{\rm{eq}}$.}   \citep[\teff{}; e.g., GJ436~b; ][]{agundez_puzzling_2014}, which coupled with its equilibrium temperature provides an estimate of the internal temperature ($T_{\rm{int}}^4 \simeq$ \teff$^4 - T_{\rm{eq}}^4$). This estimate of the internal temperature and heat flux is expected to be higher than predictions from cooling tracks (similar to \autoref{fig:coolingtrack}). Conversely, an internal temperature consistent with cooling tracks would suggest the absence of additional heating, and perhaps necessitate aerosols that provide opacity at high-altitudes and hence a larger radius. The large planet-to-stellar radius ratio and cool host star result in a large eclipse depth (\autoref{fig:emission}) making it suitable for eclipse observations with ARIEL \citep{tinetti_science_2016, helled_ariel_2022} and JWST MIRI/LRS \citep{rieke_mid-infrared_2015, wright_mid-infrared_2023}. Given that measurements of the internal temperature \citep[e.g., WASP-107~b; ][]{welbanks_high_2024} of GEMS via transmission spectroscopy are often confounded by the transit light source effect \citep{rackham_transit_2018, canas_gems_2025}, eclipse measurements may be a more suitable temperature probe.

\subsection{Sample Analysis with \texttt{MRExo}}

\begin{figure*}[ht]
\centering
\begin{tabular}{ccc}
 \includegraphics[scale=0.3, trim={0.2cm 0.2cm 0.2cm 0.2cm},clip]{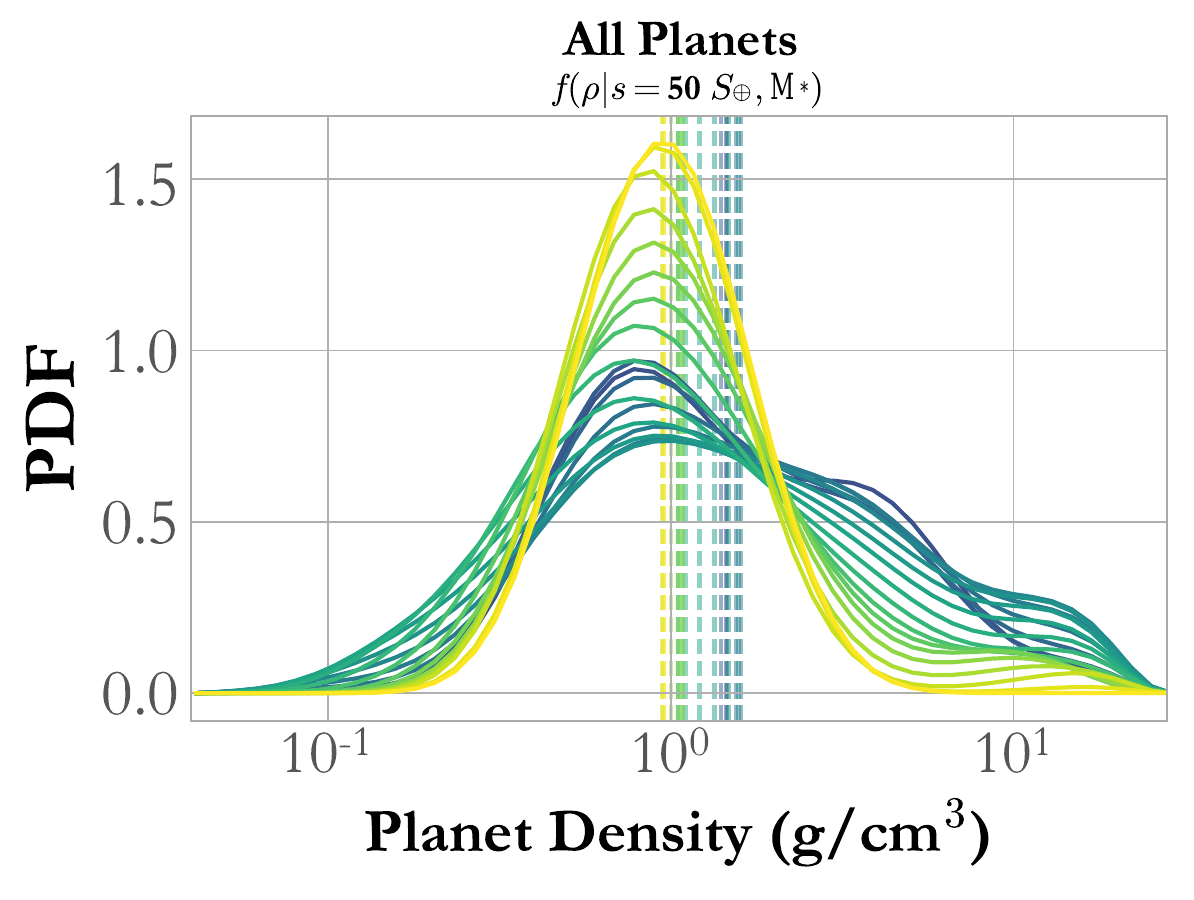} &
 \includegraphics[scale=0.3, trim={2.4cm 0.2cm 0.4cm 0.2cm},clip]{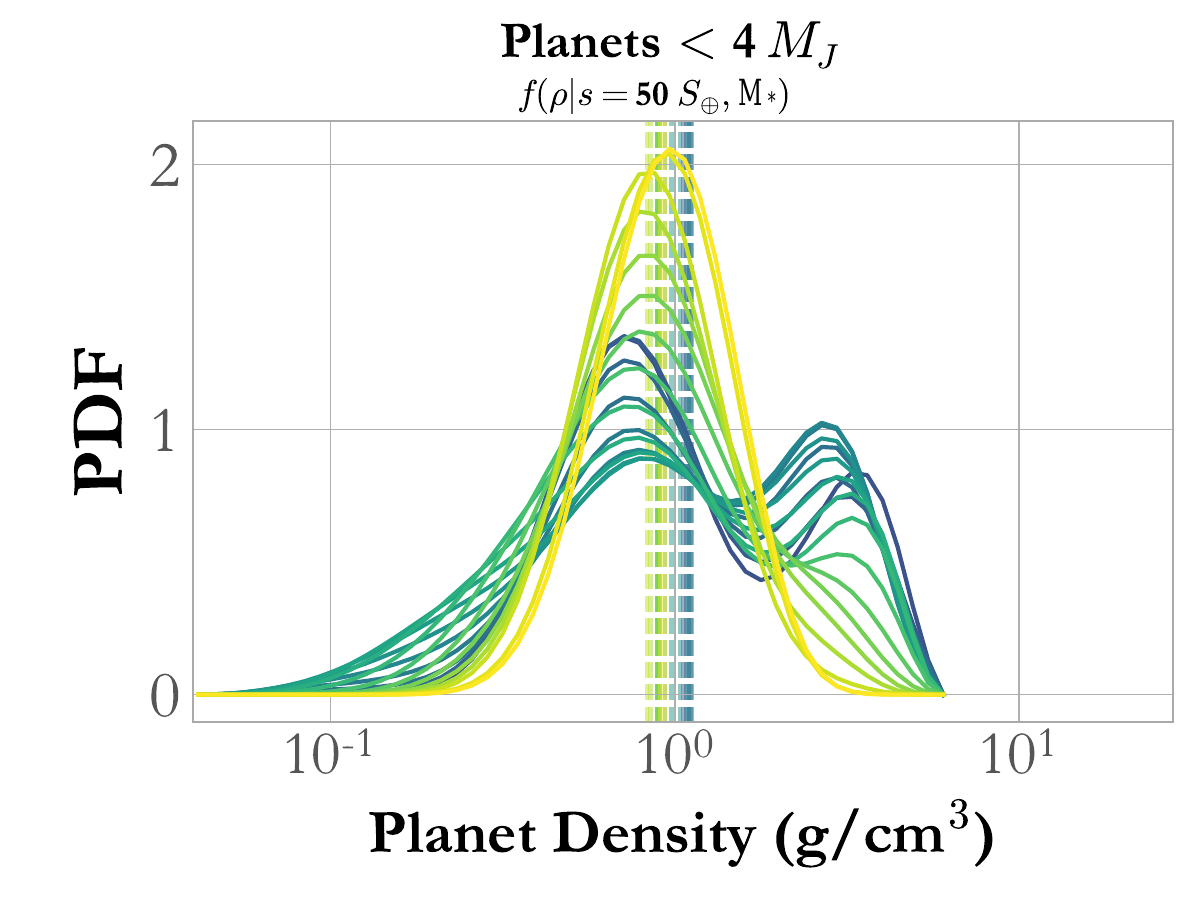} &
 \includegraphics[scale=0.3, trim={2.4cm 0.2cm 0.2cm 0.2cm},clip]{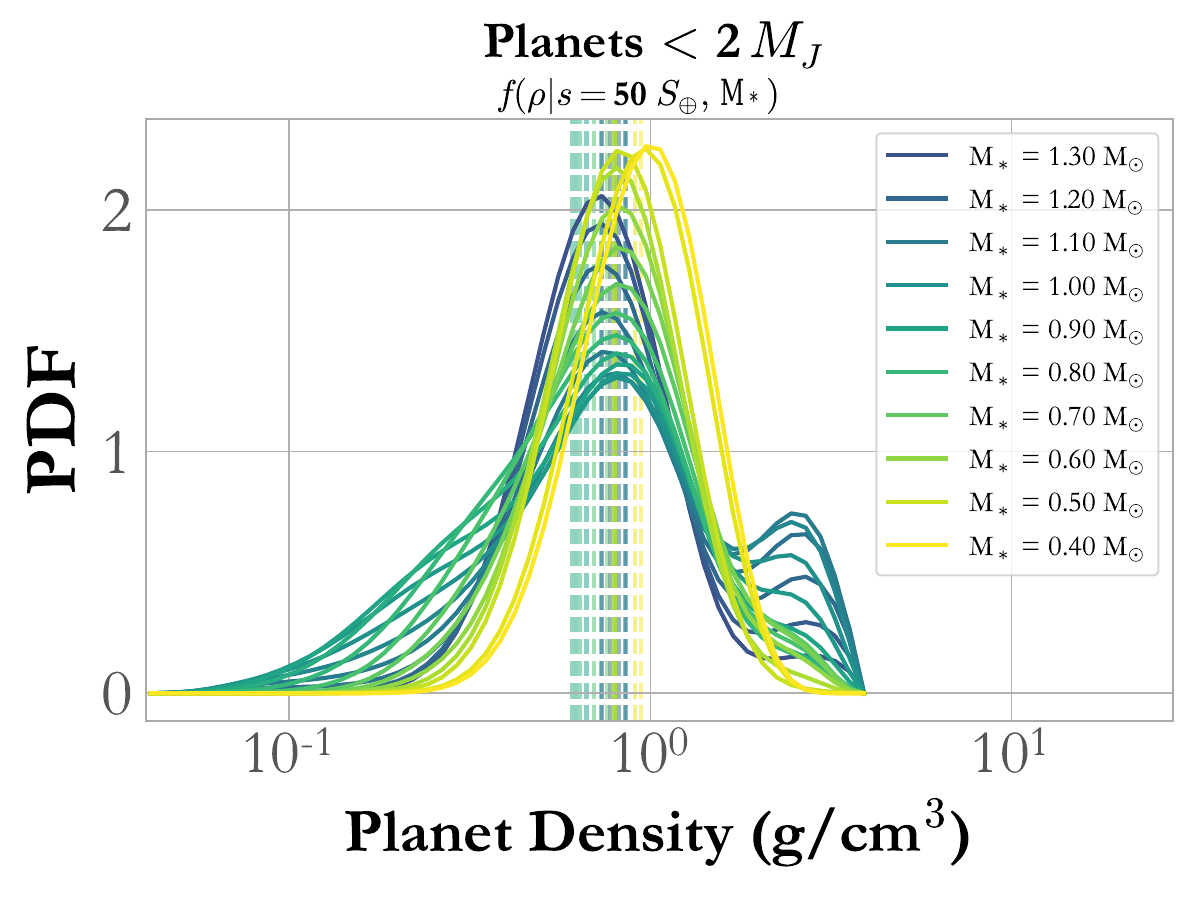}
 \end{tabular}
\caption{The 3D $f(\rho, S_p, M_*)$ distribution conditioned on different stellar masses and $S_p = 50~S_{\oplus}$ (the median insolation flux for transiting GEMS), to estimate the 1D distribution for $f(\rho~|~S_p = 50~S_{\oplus}, M_*)$. In other words, these plots show the probability distribution of bulk-densities of transiting giant planets at $50~S_{\oplus}$ across different stellar masses (shown in different colours). The left plot shows the distribution for all the giant planets in our sample. The middle and right plots show similar distributions after excluding planets $>$ 4 $M_J$ and $> 2~M_J$ from the sample. \textbf{Takeway:} We find that when excluding super-Jupiters ($> 2~M_J$), the bulk density of transiting warm Jupiters ($\sim 50~S_{\oplus}$) does not show a strong dependence on host star mass.}
\label{fig:3DConditional}
\end{figure*}

Aided by this discovery of TOI-7149~b as well as recent giant exoplanets around mid M-dwarfs --- TOI-519~b \citep{parviainen_toi-519_2021, kagetani_mass_2023}, TOI-4860~b \citep{almenara_toi-4860_2023, triaud_m_2023}, TOI-5205~b \citep{kanodia_toi-5205b_2023}, TOI-3235~b \citep{hobson_toi-3235_2023}, TOI-762~A~b \citep{hartman_toi_2024} --- we compare bulk properties of transiting giant planets across stellar masses. This is one of the primary goals of the \textit{Searching for GEMS} survey, i.e., to quantify differences between giant planets orbiting M-dwarfs and FGK stars as a means of constraining giant planet formation.

\cite{kanodia_transiting_2024} find that transiting GEMS tend to have lower masses than their FGK dwarf warm Jupiter analogues because there are fewer super-Jupiters ($\gtrsim$ 2 $M_J$) known to orbit M-dwarfs. \textbf{Once super-Jupiters are excluded from the analysis to focus on giant planets that ostensibly form through core accretion \citep[$<$ 2 -- 4 $M_J$;][]{santos_observational_2017, schlaufman_evidence_2018}, the mean mass of transiting warm Jupiters is independent of stellar mass.} This lack of dependence of warm Jupiter masses on host star mass was hypothesized to suggest that there is a minimum disk mass threshold that dominates the outcome of the core-accretion formation paradigm \citep{mizuno_formation_1980, pollack_formation_1996, laughlin_core_2004}. \cite{kanodia_transiting_2024} hypothesized that this threshold, when compared to ALMA measurements of protoplanetary disk masses, would explain the lower occurrence of transiting GEMS compared to their FGK dwarf counterparts \citep[][]{gan_occurrence_2023, bryant_occurrence_2023}, and these trends observed in their bulk properties. We emphasize that the nature of this work, i.e., comparing conditional probability distributions, is agnostic to the absolute occurrence of transiting giant planets as a function of stellar mass \citep[Section 2.2;][]{kanodia_transiting_2024}.

Following the methodology described in \cite{kanodia_searching_2024} and \cite{kanodia_transiting_2024}, we used the \textit{n}-dimensional nonparametric sample inference framework \texttt{MRExo} \citep{kanodia_beyond_2023}, which utilizes a convolution of normal and beta-density functions, to model the joint density of planets \citep{ning_predicting_2018, kanodia_mass-radius_2019}. Regions of parameter space with more planets have a higher joint probability. This $n$-dimensional joint distribution can then be marginalized to estimate the normalized conditional distribution to study the interdependence of parameters. The degree of the beta-density functions characterizes the complexity of the model required to fit the data.

We extended the analysis to additional dimensions and probed the impact of insolation flux and stellar mass on planetary bulk densities. Given that hot Jupiters with high insolation fluxes tend to be inflated due to Ohmic dissipation \citep{guillot_evolution_2002, weiss_mass_2013, thorngren_bayesian_2018}, their bulk densities cannot be trivially connected to their heavy-element (metal) content \citep{thorngren_mass-metallicity_2016, muller_towards_2022}. We therefore limited this analysis to the aforementioned sample of warm Jupiters and fit its probability density in 3 dimensions --- $M_*,~\rho_p,~S_p$ (stellar mass, bulk density, insolation). We used \textit{k}-fold cross-validation to estimate the optimum number of degrees for the beta polynomials \citep[Appendix C;][]{kanodia_beyond_2023} as 72, 41, 72 respectively.

We first fit this to all the 544 giant planets in our sample, the results of which are shown in \autoref{fig:3DConditional}. We marginalized the distribution at 50 $S_{\oplus}$ ($\sim$ 700 K, assuming 0 albedo) since that is the median insolation flux for transiting GEMS discovered so far. Thus we calculated the 1D probability distribution of planetary densities as a function of stellar mass from mid M-dwarfs to the Kraft break (0.4 -- 1.3 \solmass{}). This upper limit for stellar masses considered is to avoid the potential observational biases due to the increase in $v\sin i_{\star}$ of stars above the Kraft break  \citep[Section 5.3; ][]{kanodia_transiting_2024}. This conditional probability --- $f(\rho~|~S_p = 50~S_{\oplus}, M_*)$ --- gave us the expected planetary density for warm Jupiters receiving $50~S_{\oplus}$ from a range of host star masses.

\begin{figure}[!t] 
\centering
\includegraphics[width=0.45\textwidth]{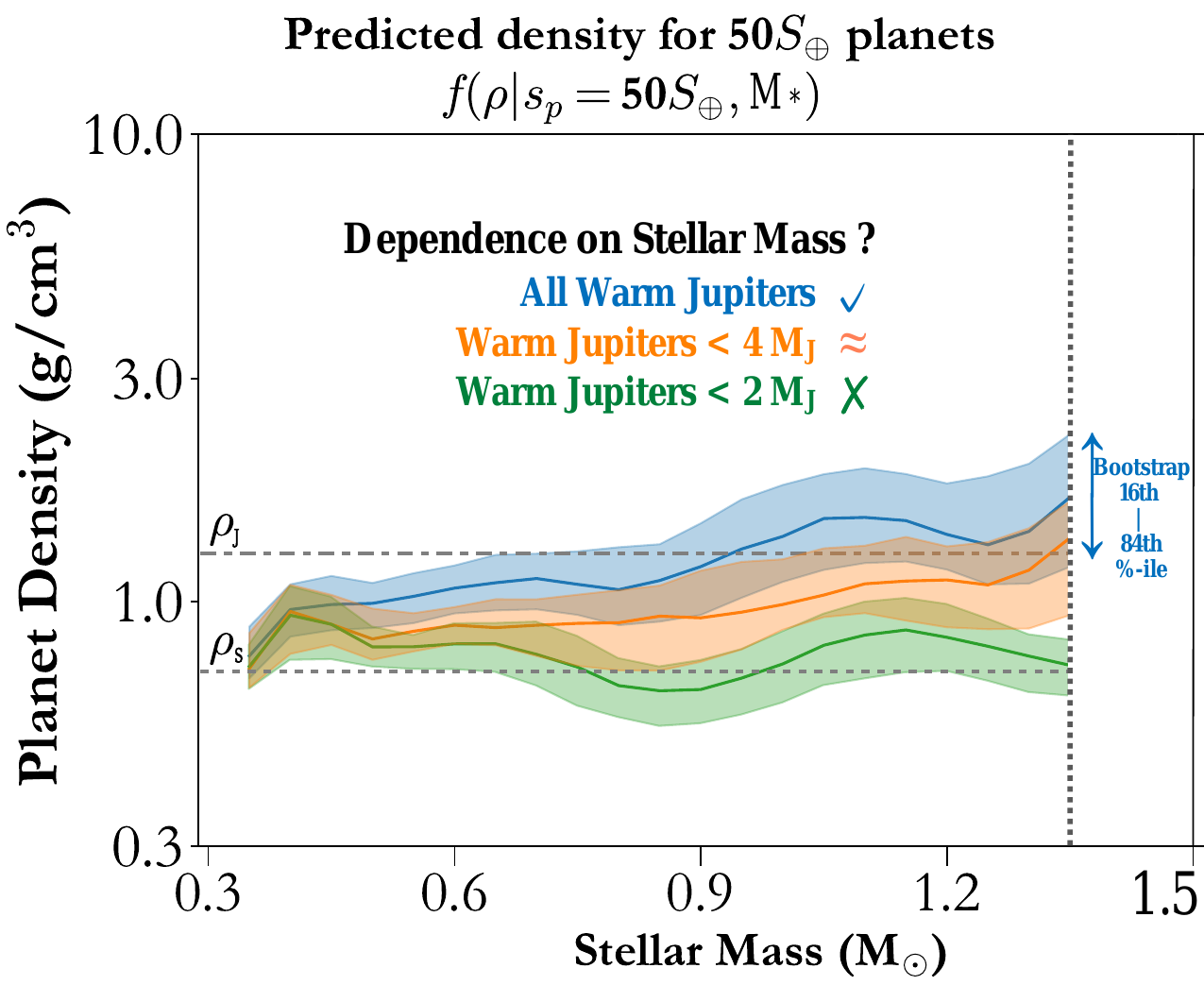}
\caption{The expectation values for the planetary bulk density as a function of stellar mass for warm Jupiters ($50~S_{\oplus}$), with the shaded regions representing the 16th~--~84th percentile uncertainty from bootstrapping the samples. The dashed horizontals line show the density of Jupiter and Saturn for reference. We do not display the stellar masses above the Kraft break due to a potential observational bias related to \vsini{} and measuring planet masses with radial velocities. \textbf{Takeaway:} For warm Jupiters unaffected by anomalous inflation \citep{thorngren_bayesian_2018}, the giant planet bulk densities offer a probe into their heavy-element content, and appear to be independent of host star mass when super Jupiters are excluded.} \label{fig:StMass_vs_Density}
\end{figure}

Next, we imposed an upper limit on the planetary mass of the input sample to investigate whether the outcomes of planet formation might be sensitive to this threshold. We test limits of 4 $M_J$, 3 $M_J$, 2.5 $M_J$, 2 $M_J$, and we show the results for the $4~M_J$ and $2~M_J$ thresholds in \autoref{fig:StMass_vs_Density}. Similar to previous results studying trends in planetary mass, we see that for planets with masses below 2 $M_J$, the planetary bulk density appears to be agnostic to the host star mass. This is consistent with the scatter plot of the input sample seen in \autoref{fig:paramspace}c, with transiting giant planets seeming to converge towards $\rho_p\sim$ 0.8 \gcmcubed{} in bulk density across the stellar mass axis.

We account for the uncertainty due to the sparsity of the sample by performing 100 bootstraps (with replacement), and show the results for the three cases in \autoref{fig:StMass_vs_Density}. We compare the mean of each of these distributions $f(\rho~|~S_p = 50~S_{\oplus}, M_*)$ using Welch's t-test \citep{welch_generalization_1947, ruxton_unequal_2006} and find consistency across the stellar mass axis. 

\begin{figure}[!t] 
\centering
\includegraphics[width=0.45\textwidth]{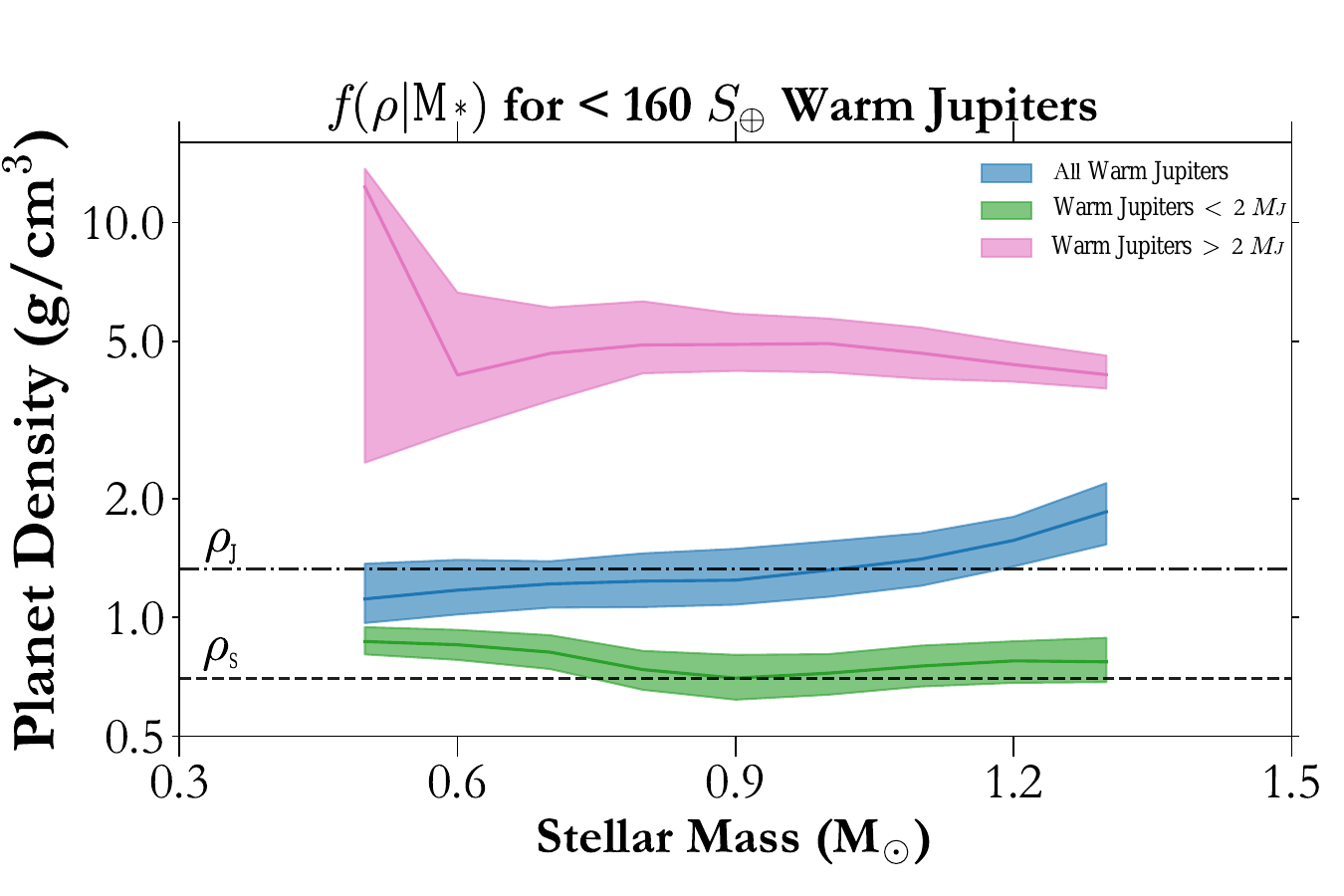}
\caption{2D analysis showing the expectation value for $f(\rho|M_*)$ as well as bootstrapped uncertainties. Similar to \autoref{fig:StMass_vs_Density}, we see that the sample of warm Jupiters $<$ 160 $S_{\oplus}$ (blue; T$_{eq}$ $\sim$ 1000 K) have bulk densities that appear to show a stellar mass dependency. This disappears when considering a sample of planets below $<$ 2 $M_J$ (green), suggesting different formation or evolution mechanisms across these two subsamples with contrasting stellar mass dependence. In pink we also show the super-Jupiters $>$ 2 $M_J$. The large bootstrap uncertainty in this for the M-dwarfs ($< 0.6$ \solmass) is driven by the number of super-Jupiters around M-dwarfs ($<$~5). Since the relative number of super-Jupiters to Jupiters increases with increasing stellar mass (\autoref{fig:paramspace}) the bulk density of the combined sample (blue curve) shows a stellar mass dependence, whereas intrinsically the $< 2$ $M_J$ do not. } \label{fig:2DStMass_vs_Density}
\end{figure}

Instead of marginalizing for a single insolation flux, we also repeat this analysis on warm Jupiters with $S_p<$ 160 $S_{\oplus}$ (shown in solid in \autoref{fig:paramspace}c), and find a similar result, with $f(\rho|M_*)$ being constant for the $< 2 M_J$ sample, but varying with stellar mass for the combined sample due to the varying occurrence super-Jupiters and Jupiters as a function of stellar mass (\autoref{fig:2DStMass_vs_Density}). We add the caveat that while planet density is a reasonable proxy for bulk metallicity, it might not necessarily be so if there are systematic differences in the ages, formation and cooling history, or atmospheric composition across the stellar mass axis for giant planets, as suggested by \cite{muller_bulk_2024} and being investigated with a JWST survey of GEMS \citep{kanodia_red_2023}.

\subsection{Implications of lack of dependence of bulk density on stellar mass}

We note that under the core accretion paradigm, upon initiation of the runaway processes --- typically assumed to be between 20 -- 100 \earthmass{} \citep{helled_mass_2023}, depending on the grain size, surface density, gas opacity, and protoplanet gas-to-metal ratio  --- the protoplanets are expected to exponentially accrete gas from their host disk. This process is only possible until the gas flow to the planet is quenched due to a gap opening up in the disk \citep{lin_tidal_1993, bryden_tidally_1999}. Furthermore, given the effect of electron degeneracy pressure beyond Saturn mass \citep[$\sim 0.3 M_J$;][]{saumon_theory_1996}, objects in our sample can have masses ranging from Saturn to the Hydrogen-burning mass upper limit while exhibiting a radius comparable to Jupiter. Thus, for the subsample under consideration ($<2 M_J$),  densities ranging from $\rho_S$~--~$2 \rho_J$ (0.7 -- 2.6 \gcmcubed{}), i.e, a factor of $\sim 4$ are possible. Yet, we see that the densities of warm Jupiters seem to converge around 0.8~\gcmcubed. 

While this can be interpreted as being consistent with the hypothesis proposed in our previous work \citep[see Figure 8,][]{kanodia_transiting_2024}, in which a minimum disk mass threshold dictates the final outcome of giant planet bulk densities (as well as their inverse dependent occurrence with stellar mass), we do not account for the impact of migration and evolution during and post formation. Such processes may play a significant role but investigating these effects is beyond the scope of this work. 

We also note that the absence of a stellar mass dependence for the warm Jupiters with masses $<2 M_J$ could potentially indicate a different formation mechanism for planets below this threshold. This has been previously suggested by \cite{santos_observational_2017}, \cite{schlaufman_evidence_2018} and \cite{narang_properties_2018} who found a mass upper limit for giant planets around 4 -- 10 $M_J$, and interpreted this boundary as the threshold where the dominant formation mechanism transitions from core accretion to gravitational instability \citep{boss_giant_1997}.

Overall this suggests that these GEMS ($< 2$ $M_J$) --- including the ones around mid-to-late M-dwarfs such as TOI-7149~b (this work), TOI-5205~b \citep{kanodia_toi-5205b_2023}, TOI-3235~b \citep{hobson_toi-3235_2023}, which are a challenge to explain with core accretion due to insufficient disk masses --- still likely form in this conventional manner in protoplanetary disks. However, this likely takes place in disks that are very massive for the stars they orbit in order to satisfy the minimum disk mass threshold mentioned above.

\section{Summary}\label{sec:conclusion}
As part of the \textit{Searching for GEMS} survey, we describe the discovery of TOI-7149~b, a low-density planet extending the transiting GEMS sample to lower-mass host stars with a large $\sim$ 12\% transit depth. The host star is a fully convective M-dwarf, and one of the coldest to host a gas giant planet. With ground-based RVs from HPF and transits from Palomar, RBO, TMMT and LCRO, we obtain a precise planetary mass, radius and planetary density measurement. TOI-7149~b has an inflated radius that cannot be explained by its mass and insolation flux, and can be probed by constraining its internal temperature using emission spectroscopy. We use the statistical framework \texttt{MRExo} and extend our analysis from previous work to analyze the sample of transiting warm Jupiters and find a stellar mass dependency for their bulk densities when considering the entire sample of transiting warm Jupiters. This dependency vanishes when focussing on a subset of warm Jupiters with planetary masses $< 2 M_J$, suggesting (i) different formation pathways for Jupiters below $\sim$ 2 $M_J$, and (ii) for the $< 2 M_J$, the formation mechanism seems to be agnostic to the host star mass. This independence on stellar mass could hint at a minimum threshold disk mass required for giant planet formation, which when contextualized with ALMA disk dust mass measurements in future work could explain the varying occurrence of these objects with stellar mass, as well as the above mentioned trend.

\section{Acknowledgements}

We thank the anonymous referee for the valuable feedback which has improved the quality of this manuscript.

SK acknowledges research support from Carnegie Institution of Science through the Carnegie Fellowship. SK would like to thank Peter Gao for help with the computing resources that enabled running some of these memory intensive \texttt{MRExo} analyses. SK acknowledges feedback from Simon M\"{u}ller regarding the discussion in this paper.

We acknowledge support from U.S. Contributions to Ariel Preparatory
Science NASA grant 80NSSC25K0184.

% HET/HPF, also needs footnote
These results are based on observations obtained with the Habitable-zone Planet Finder Spectrograph on the HET. We acknowledge support from NSF grants AST-1006676, AST-1126413, AST-1310885, AST-1310875,  ATI-2009889, ATI-2009554, ATI-2009982, AST-2108512, AST-2108801 and the NASA Astrobiology Institute (NNA09DA76A) in the pursuit of precision radial velocities in the NIR. The HPF team also acknowledges support from the Heising-Simons Foundation via grant 2017-0494. 

The Hobby-Eberly Telescope is a joint project of the University of Texas at Austin, the Pennsylvania State University, Ludwig-Maximilians-Universität München, and Georg-August Universität Gottingen. The HET is named in honor of its principal benefactors, William P. Hobby and Robert E. Eberly. The HET collaboration acknowledges the support and resources from the Texas Advanced Computing Center. We thank the Resident astronomers and Telescope Operators at the HET for the skillful execution of our observations with HPF. We would like to acknowledge that the HET is built on Indigenous land. Moreover, we would like to acknowledge and pay our respects to the Carrizo \& Comecrudo, Coahuiltecan, Caddo, Tonkawa, Comanche, Lipan Apache, Alabama-Coushatta, Kickapoo, Tigua Pueblo, and all the American Indian and Indigenous Peoples and communities who have been or have become a part of these lands and territories in Texas, here on Turtle Island.

%WIYN acknowledgment
WIYN is a joint facility of the University of Wisconsin-Madison, Indiana University, NSF's NOIRLab, the Pennsylvania State University, Purdue University, University of California-Irvine, and the University of Missouri. 

%Land acknowledgment
The authors are honored to be permitted to conduct astronomical research on Iolkam Du'ag (Kitt Peak), a mountain with particular significance to the Tohono O'odham. Data presented herein were obtained at the WIYN Observatory from telescope time allocated to NN-EXPLORE through the scientific partnership of NASA, the NSF, and NOIRLab.

% NESSI
Some of the observations in this paper made use of the NN-EXPLORE Exoplanet and Stellar Speckle Imager (NESSI). NESSI was funded by the NASA Exoplanet Exploration Program and the NASA Ames Research Center. NESSI was built at the Ames Research Center by Steve B. Howell, Nic Scott, Elliott P. Horch, and Emmett Quigley.

% Gaia
This work has made use of data from the European Space Agency (ESA) mission Gaia (\url{https://www.cosmos.esa.int/gaia}), processed by the Gaia Data Processing and Analysis Consortium (DPAC, \url{https://www.cosmos.esa.int/web/gaia/dpac/consortium}). Funding for the DPAC has been provided by national institutions, in particular the institutions participating in the Gaia Multilateral Agreement.

% ACI
Computations for this research were performed on the Pennsylvania State University’s Institute for Computational and Data Sciences Advanced CyberInfrastructure (ICDS-ACI).  This content is solely the responsibility of the authors and does not necessarily represent the views of the Institute for Computational and Data Sciences.

% CEHW 
The Center for Exoplanets and Habitable Worlds is supported by the Pennsylvania State University, the Eberly College of Science, and the Pennsylvania Space Grant Consortium. 

% MAST
Some of the data presented in this paper were obtained from MAST at STScI. Specifically, the tglc data products are available as a High Level Science Product \citep{10.17909/610m-9474}. Support for MAST for non-HST data is provided by the NASA Office of Space Science via grant NNX09AF08G and by other grants and contracts.

% Kepler/TESS
This work includes data collected by the TESS mission, which are publicly available from MAST. Funding for the TESS mission is provided by the NASA Science Mission directorate. 

% NASA Exoplanet Archive, ADS, 2MASS
This research made use of the (i) NASA Exoplanet Archive, which is operated by Caltech, under contract with NASA under the Exoplanet Exploration Program, (ii) SIMBAD database, operated at CDS, Strasbourg, France, (iii) NASA's Astrophysics Data System Bibliographic Services, and (iv) data from 2MASS, a joint project of the University of Massachusetts and IPAC at Caltech, funded by NASA and the NSF.

%ADS
This research has made use of the SIMBAD database, operated at CDS, Strasbourg, France, 
and NASA's Astrophysics Data System Bibliographic Services.

% Exofop 
This research has made use of the Exoplanet Follow-up Observation Program website, which is operated by the California Institute of Technology, under contract with the National Aeronautics and Space Administration under the Exoplanet Exploration Program

% Caleb
CIC acknowledges support by an appointment to the NASA Postdoctoral Program at the Goddard Space Flight Center, administered by ORAU through a contract with NASA.

% SK would like to acknowledge Theodora and Rafa for help with this project.
The research was carried out, in part, at the Jet Propulsion Laboratory, California Institute of Technology, under a contract with the National Aeronautics and Space Administration (80NM0018D0004)

This research made use of \textsf{exoplanet} \citep{foreman-mackey_exoplanet-devexoplanet_2021, foreman-mackey_exoplanet_2021} and its
dependencies \citep{agol_analytic_2020, foreman-mackey_fast_2017, foreman-mackey_scalable_2018, kumar_arviz_2019, robitaille_astropy:_2013, astropy_collaboration_astropy_2018, kipping_efficient_2013, luger_starry_2019, the_theano_development_team_theano_2016, salvatier_probabilistic_2016}

\facilities{\gaia{}, HET (HPF),  WIYN 3.5 m (NESSI), RBO, Palomar (WIRC), \tess{}, Exoplanet Archive, TMMT, LCRO}
\software{
\texttt{ArviZ} \citep{kumar_arviz_2019}, 
AstroImageJ \citep{collins_astroimagej_2017}, 
\texttt{astroquery} \citep{ginsburg_astroquery_2019}, 
\texttt{astropy} \citep{robitaille_astropy:_2013, astropy_collaboration_astropy_2018},
\texttt{barycorrpy} \citep{kanodia_python_2018}, 
\texttt{celerite2} \citep{foreman-mackey_fast_2017, foreman-mackey_scalable_2018},
\texttt{exoplanet} \citep{foreman-mackey_exoplanet-devexoplanet_2021, foreman-mackey_exoplanet_2021},
\texttt{EXOFASTv2} \citep{eastman_exofastv2_2019}, 
\texttt{HPF-SERVAL} \citep{Stefansson2020},
\texttt{HPF-SpecMatch} \citep{Stefansson2020},
\texttt{HxRGproc} \citep{ninan_habitable-zone_2018},
\texttt{GALPY} \citep{bovy_galpy_2015},
\texttt{ipython} \citep{perez_ipython:_2007},
\texttt{lightkurve} \citep{lightkurve_collaboration_lightkurve_2018},
\texttt{matplotlib} \citep{hunter_matplotlib:_2007},
\texttt{MRExo} \citep{kanodia_mass-radius_2019, kanodia_beyond_2023},
\texttt{numpy} \citep{oliphant_numpy:_2006},
\texttt{pandas} \citep{mckinney-proc-scipy-2010},
\texttt{planetsynth} \citep{muller_synthetic_2021},
\texttt{PyMC3} \citep{salvatier_probabilistic_2016},
\texttt{scipy} \citep{oliphant_python_2007, virtanen_scipy_2020},
\texttt{starry} \citep{luger_starry_2019, agol_analytic_2020},
\texttt{tglc} \citep{han_tess-gaia_2023},
\texttt{tessminer} (Glusman et al. in prep.),
\texttt{Theano} \citep{the_theano_development_team_theano_2016},
\texttt{wotan} \citep{hippke_wotan_2019}
}

\bibliography{MyLibrary, ManualReferences}
\bibliographystyle{aasjournal}

\end{document}